\documentclass{aa}  
\usepackage{natbib}
\usepackage{graphicx}
\usepackage{txfonts}
\usepackage{hyperref}
\usepackage{booktabs}
\newcommand{\numax}{\mbox{$\nu_{\rm max}$}}
\newcommand{\deltanu}{\mbox{$\langle \Delta\nu \rangle$}}
\newcommand{\calnum}{\mathcal{C}_{\nu_{\rm max}}}
\newcommand{\caldnu}{\mathcal{C}_{\langle \Delta\nu \rangle}}

\newcommand{\ogaia}{\mathcal{O}_{\rm Gaia}}

\usepackage[inline]{enumitem}
\usepackage{gensymb}

\begin{document} 
  \title{New light on the \textit{Gaia} DR2 parallax zero-point: influence of the asteroseismic approach, in and beyond the \textit{Kepler} field}
  \titlerunning{\textit{Gaia} DR2 parallax zero-point: influence of the asteroseismic approach}

  %\subtitle{}
  \author{S. Khan\inst{1, 2}
        \and
        A. Miglio\inst{1, 2}
        \and
        B. Mosser\inst{3}
        \and
        F. Arenou\inst{4}
        \and
        K. Belkacem\inst{3}
        \and
        A. G. A. Brown\inst{5}
        \and
        D. Katz\inst{4}
        \and
        L. Casagrande\inst{6, 7}
        \and
        W. J. Chaplin\inst{1, 2}
        \and
        G. R. Davies\inst{1, 2}
        \and
        B. M. Rendle\inst{1, 2}
        \and
        T. S. Rodrigues\inst{8}
        \and
        D. Bossini\inst{8}
        \and
        T. Cantat-Gaudin\inst{9}
        \and
        Y. P. Elsworth\inst{1, 2}
        \and
        L. Girardi\inst{8}
        \and
        T. S. H. North\inst{1}
        \and
        A. Vallenari\inst{8}
                }

  \institute{School of Physics and Astronomy, University of Birmingham,
              Edgbaston, Birmingham, B15 2TT, UK\\
              \email{sxk1008@bham.ac.uk}
         \and
             Stellar Astrophysics Centre, Department of Physics and Astronomy, Aarhus University, Ny Munkegade 120, DK-8000 Aarhus C, Denmark
         \and
             LESIA, Observatoire de Paris, PSL Research University, CNRS, Sorbonne Universit\'e, Universit\'e Paris Diderot, 92195 Meudon, France
         \and
             GEPI, Observatoire de Paris, PSL Research University, CNRS, 92195 Meudon, France
         \and
             Leiden Observatory, Leiden University, Niels Bohrweg 2, 2333, CA, Leiden, The Netherlands
         \and
             Research School of Astronomy and Astrophysics, Mount Stromlo Observatory, The Australian National University, ACT 2611, Australia
         \and
             ARC Centre of Excellence for All Sky Astrophysics in 3 Dimensions (ASTRO 3D) 
         \and
             Osservatorio Astronomico di Padova, INAF, Vicolo dell'Osservatorio 5, I-35122 Padova, Italy
         \and
             Institut de Ciències del Cosmos, Universitat de Barcelona (IEEC-UB), Martí i Franquès 1, 08028, Barcelona, Spain
            }

  \date{Received September 15, 1996; accepted March 16, 1997}

  \abstract{The importance of studying the \textit{Gaia} DR2 parallax zero-point by external means was underlined by \citet{Lindegren2018}, and initiated by several works making use of Cepheids, eclipsing binaries, and asteroseismology. Despite a very efficient elimination of basic-angle variations, a small fluctuation remains and shows up as a small offset in the \textit{Gaia} DR2 parallaxes. By combining astrometric, asteroseismic, spectroscopic, and photometric constraints, we undertake a new analysis of the \textit{Gaia} parallax offset for nearly 3000 red-giant branch (RGB) and 2200 red clump (RC) stars observed by \textit{Kepler}, as well as about 500 and 700 red giants (both RGB and RC) selected by the K2 Galactic Archaeology Program in campaigns 3 and 6. Engaging into a thorough comparison of the astrometric and asteroseismic parallaxes, we are able to highlight the influence of the asteroseismic method, and measure parallax offsets in the \textit{Kepler} field that are compatible with independent estimates from literature and open clusters. Moreover, adding the K2 fields to our investigation allows us to retrieve a clear illustration of the positional dependence of the zero-point, in general agreement with the information provided by quasars. Lastly, we initiate a two-step methodology to make progress in the simultaneous calibration of the asteroseismic scaling relations and of the \textit{Gaia} DR2 parallax offset, which will greatly benefit from the gain in precision with the third Data Release of \textit{Gaia}.}


\maketitle

%________________________________________________________________

\section{Introduction}
\label{sec:intro}

Masses and radii of solar-like oscillating stars can be estimated from the global asteroseismic observables that characterise their oscillation spectra, namely the average large frequency separation ($\deltanu$) and the frequency corresponding to the maximum observed oscillation power ($\numax$). The large frequency spacing is predicted by theory to approximately scale as the square root of the mean density of the star \citep[see, e.g.,][]{Vandakurov1967,Tassoul1980}:
\begin{align}
    \deltanu \propto \sqrt{\langle \rho \rangle} \propto \sqrt{\frac{M}{R^3}} \, ,
    \label{eq:Dnu}
\end{align}
where $M$ and $R$ are the stellar mass and radius, respectively. 
The frequency of maximum power follows to good approximation a proportional relation with 
the acoustic cut-off frequency, and can provide a direct measure of the surface gravity ($g$) when the effective temperature ($T_{\rm eff}$) is known \citep[see, e.g.,][]{Brown1991,Kjeldsen1995,Belkacem2011}:
\begin{align}
    \numax \propto \frac{g}{\sqrt{T_{\rm eff}}} \propto  \frac{M}{R^2\sqrt{T_{\rm eff}}} \, .
    \label{eq:numax}
\end{align}
Equations \ref{eq:Dnu} and \ref{eq:numax} imply that if $\deltanu$ and $\numax$ are available, together with an independent estimate of $T_{\rm eff}$, a ``direct'' estimation of the stellar mass and radius is possible. This direct method is particularly attractive because it provides, in principle, estimates that are independent of stellar models. 
Alternatively, one may also use $\deltanu$ and $\numax$ as inputs to a grid-based estimation of the stellar properties, matching the observations to stellar evolutionary tracks --- either using the scalings at face value or stellar pulsation calculations to obtain $\deltanu$ \citep[e.g.][]{Stello2009,Basu2010,Gai2011,Rodrigues2017}. Whether it be with the direct or the grid-based approach, a plethora of studies have compared asteroseismic measurements of radii (or distances) with independent ones, such as clusters \citep{Miglio2012,Miglio2016,Stello2016,Handberg2017}, interferometry \citep{Huber2012}, eclipsing binaries \citep{Gaulme2016,Brogaard2016,Brogaard2018}, and astrometry \citep{SilvaAguirre2012,DeRidder2016,Davies2017,Huber2017,Sahlholdt2018,Zinn2018}. All these works indicated that stellar radius estimates from asteroseismology are accurate to within a few percent.

On the astrometric side, before the \textit{Gaia} data, 
the asteroseismic distances --- arising from the combination of seismic constraints with effective temperature and apparent photometric magnitudes --- of stars in the solar neighbourhood had only been compared a posteriori with \textsc{Hipparcos} values, 
with limitations due to the \textsc{Hipparcos} uncertainties being large for most of the \textit{Kepler} and CoRoT targets \citep{Miglio2012,SilvaAguirre2012,Lagarde2015}. 
The announcement of the first \textit{Gaia} Data Release opened the gates to the \textit{Gaia} era \citep{GaiaCollaboration2016,GaiaCollaboration2016a}. 
Parallaxes and proper motions were available for the 2 million brightest sources in \textit{Gaia} DR1, as part of the \textit{Tycho-Gaia} Astrometric Solution \citep[TGAS;][]{Lindegren2016}. 
As the TGAS parallaxes considerably improved the \textsc{Hipparcos} values, a new comparison between astrometric and asteroseismic parallaxes was appropriate. Some works took the path of the model-independent method, i.e. using asteroseismic distances based on the use of the raw scaling relations. 
Using assumptions about the luminosity of the red clump, \citet{Davies2017} found the TGAS sample to overestimate the distance, with a median parallax offset of $-0.1$ mas. 
For 2200 \textit{Kepler} stars, from the main sequence to the red-giant branch, \citet{Huber2017} obtained a qualitative agreement, especially if they adopted a hotter effective temperature scale for dwarfs and subgiants. The latter suggestion was corroborated by \citet{Sahlholdt2018}. In contrast, \citet{DeRidder2016} used seismic modelling methods to analyse two samples of stars observed by \textit{Kepler}: 22 nearby dwarfs and subgiants showing an excellent overall correspondence; and 938 red giants for which the TGAS parallaxes were significantly smaller than the seismic ones. Given the different seismic approaches and the various outcomes, the situation as regards to the \textit{Gaia} DR1 parallax offset, as probed by asteroseismology, was left unclear. 

The second Data Release of \textit{Gaia} was published on April 25th, 2018 \citep{GaiaCollaboration2018}, based on the data collected during the first 22 months of the nominal mission lifetime \citep{GaiaCollaboration2016a}. \textit{Gaia} DR2 represents a major advance with respect to the first intermediate \textit{Gaia} Data Release, containing parallaxes and proper motions for over 1.3 billion sources. 
Unlike the TGAS, the \textit{Gaia} DR2 astrometric solution does not incorporate any information from \textsc{Hipparcos} and \textit{Tycho-2}. 
However, with less than two years of observations and preliminary calibrations, a few weaknesses in the quality of the astrometric data remain, and were identified by \citet{Arenou2018} and \citet{Lindegren2018}. Among these caveats, the latter study underlined the importance of investigating the parallax zero-point by external means, 
and did so through the use of quasars which are a quasi-ideal means in this matter: negligibly small parallaxes, large number, and availability over most of the celestial sphere. 
A global zero-point of about $-30 \ \mu$as was found by \citet{Lindegren2018}, in the sense that \textit{Gaia} parallaxes are too small, with variations --- in the order of several tens of $\mu$as --- depending on a given combination of magnitude, colour, and position. Quasars have their own specific properties, such as their faintness and blue colour, which should be kept in mind when interpreting these results. For this reason, a direct correction of individual parallaxes from the global parallax zero-point is discouraged \citep{Arenou2018}.

In this context, several works have confirmed the existence of a parallax offset by independent means. The study of 50 Cepheids by \citet{Riess2018} suggested a zero-point offset of $-46 \pm 13 \ \mu$as. 
\citet{Stassun2018} presented evidence of a systematic offset of about $-82 \pm 33 \ \mu$as with 89 eclipsing binaries. And, finally, \citet{Zinn2018} inferred a systematic error of $-52.8 \pm 2.4$ (statistical) $\pm 1$ (systematic) $\mu$as for 3500 first-ascent giants in the \textit{Kepler} field, using asteroseismic and spectroscopic constraints from \citet{Pinsonneault2018} who used model-predicted corrections to the $\deltanu$ scaling relation. Very little difference was found with 2500 red-clump stars: $-50.2 \pm 2.5$ (statistical) $\pm 1$ (systematic) $\mu$as, which is expected from the astrometric point of view since \textit{Gaia}, unlike seismology, does not make any distinction between shell-hydrogen and core-helium burning stars.

These various outcomes demonstrate the need to independently solve the parallax zero-point within the framework of an analysis having its own specificities, i.e. magnitude, colour, and spatial distributions. In the case of asteroseismology, the findings of a comparison with \textit{Gaia} DR2 cannot be dissociated from the seismic method employed. With this in mind, we engage into a thorough investigation of the \textit{Gaia} DR2 parallax offset in the \textit{Kepler} field, by taking an incremental approach --- starting with the scaling relations taken at face value and gradually working towards a 
Bayesian estimation of stellar properties using a grid of models (Sect. \ref{sec:kepler}). Also, looking at the broader picture and considering two fields of the re-purposed \textit{Kepler} mission, K2, allows us to investigate the positional dependence of the zero-point (Sect. \ref{sec:position}). Lastly, \textit{Gaia} DR2 offers scope for development in the calibration of the scaling relations, and we initiate a two-step methodology allowing us to constrain the \textit{Gaia} DR2 offset at the same time (Sect. \ref{sec:calibration}).

%__________________________________________________________________

\section{Observational framework}
\label{sec:obs}

\subsection{Description of the datasets}
\label{sec:datasets}

One part of our sample consists of red-giant stars observed by \textit{Kepler} and with APOGEE spectra available \citep[APOKASC collaboration;][]{Abolfathi2018}. From the initial list of stars, we select those that are classified as RGBs and RCs (including secondary clump stars as well) using the method by \citet{Elsworth2017}. We consider the global asteroseismic parameters $\deltanu$ and $\numax$. We use the frequency of maximum oscillation power, $\numax$, from \citet{Mosser2011}. Two methods for providing relevant estimates of $\deltanu$ are discussed in Sect. \ref{sec:Dnu}. 
We also make use of the spectroscopically measured effective temperature $T_{\rm eff}$, surface gravity $\log g$ (calibrated against asteroseismic surface gravities), and of constraints on the photospheric chemical composition [Fe/H] and [$\alpha$/Fe] from SDSS DR14, as determined by the APOGEE Pipeline \citep{Abolfathi2018}. This leads us to 3159 RGB stars and 2361 RC stars in the \textit{Kepler} field.

Our \textit{Kepler} subsample is then complemented with red giants selected by the K2 Galactic Archaeology Program \citep[GAP;][]{Howell2014,Stello2015,Stello2017} in campaigns 3 (south Galactic cap) and 6 (north Galactic cap), that have SkyMapper photometric constraints \citep{Casagrande2019}. We make use of the asteroseismic analysis method from \citet{Mosser2011} for the vast majority and from \citet{Elsworth2017} for a very small fraction of stars ($\sim 5 \, \%$), and we refer the reader to Rendle et al. (in prep.) for details about additional tests of the seismic results' reliability. 
For $T_{\rm eff}$ and [Fe/H], we use the photometric estimates originating from the SkyMapper survey \citep{Casagrande2019}, and $\log g$ comes from asteroseismically-derived estimates. This K2 subsample falls into two parts: 505 and 723 red giants in C3 and C6, respectively. 

For the full sample, stellar masses and extinctions are inferred using the Bayesian tool \textsc{param} \citep{Rodrigues2014,Rodrigues2017}. Asteroseismic constraints $\deltanu$ and $\numax$ are included in the modelling procedure in a self-consistent manner, whereby $\deltanu$ is calculated from a linear fitting of the individual radial-mode frequencies of the models in the grid. \textsc{param} also requires photometry, and uses its own set of bolometric corrections \citep[described at length in][]{Girardi2002}, to estimate distances and extinctions. In addition, we calculate extinctions via the \citet{Green2015} dust map and the RJCE method \citep{Majewski2011} for comparison purposes. The bolometric corrections are derived using the code written by \citet{Casagrande2014a,Casagrande2018a,Casagrande2018}, taking $T_{\rm eff}$, $\log g$, and [Fe/H] as input parameters. The second Data Release of \textit{Gaia} \citep{GaiaCollaboration2016a,GaiaCollaboration2018} then provides us with astrometric and photometric constraints: parallaxes (using the external parallax uncertainty as described by Lindegren et al. in their overview of \textit{Gaia} DR2 astrometry\footnote{\url{https://www.cosmos.esa.int/web/gaia/dr2-known-issues}}, and made available by the \textit{Gaia} team at the GEPI, Observatoire de Paris\footnote{\url{https://gaia.obspm.fr/tap-server/tap}}), $G$-band magnitudes --- which are corrected following \citet{Casagrande2018a}, i.e. $G_{\rm corr}=0.0505+0.9966 \, G$ --- and $G_{\rm BP}-G_{\rm RP}$ colour indices. The 2MASS \citep{Skrutskie2006} $K$-band photometry is used as well.

\subsection{Consistency in the definition of $\deltanu$}
\label{sec:Dnu}

\begin{figure}
   \centering
   \includegraphics[width=\hsize]{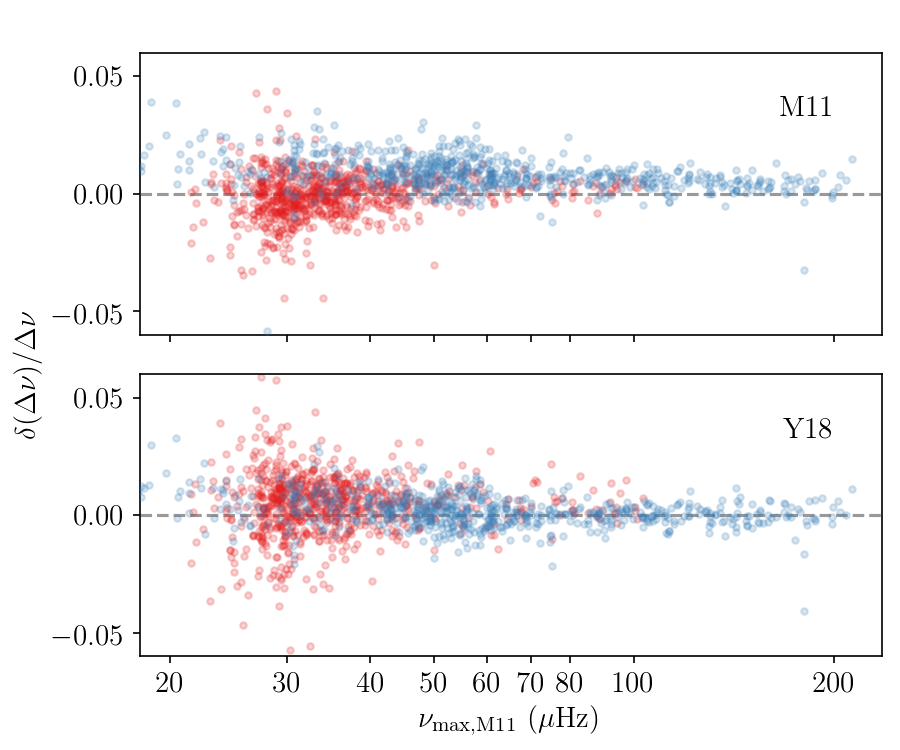}
   \caption{Relative difference in $\deltanu$, $\delta(\Delta \nu) / \Delta \nu = (\Delta \nu_{\rm other}- \Delta \nu_{\rm D16})/ \Delta \nu_{\rm D16}$, between individual frequencies following \citet{Davies2016a} (D16) and another method, as a function of $\numax$ as estimated by \citet{Mosser2011}. The comparison is done with \citet{Mosser2011} (M11; top) and \citet{Yu2018} (Y18; bottom). RGB and RC stars are in blue and red, respectively. Here, $\Delta \nu$ is used, instead of $\deltanu$, to simplify the notation.}
   \label{fig:Dnu_comp}
\end{figure}

For the \textit{Kepler} field (Sect. \ref{sec:kepler}), we explore different seismic methods, which have to be matched with a consistent definition of $\deltanu$. To use $\deltanu$ in the scaling relations, one would want to adopt a measure which is as close as possible to the asymptotic limit (on which the scaling is based). This implies, e.g., correcting for acoustic glitches \citep[regions of sharp sound-speed variation in the stellar interior related to a rapid change in the chemical composition, the ionisation of major chemical elements, or the transition from radiative to convective energy transport; see, e.g.,][]{Miglio2010,Vrard2015}. In that case, the $\deltanu$ measured by \citet{Mosser2011} is appropriate since their method mitigates the perturbation on $\deltanu$ due to glitches. On the other hand, one could abandon scalings and use $\deltanu$ from models which can, e.g., be based on individual frequencies as in \textsc{param}, which also takes into account departures from homology (regarding the assumption of scaling with density, i.e. Eq. (\ref{eq:Dnu})). Then, it is more adequate to combine \textsc{param} with $\deltanu$ estimates from individual radial-mode frequencies. While the latter are currently available only for a small subset (697 RGB and 783 RC stars), following the approach presented in \citet{Davies2016a}, we notice a qualitative agreement with $\deltanu$ from \citet{Yu2018} which are also derived from the frequencies. On the contrary, the $\deltanu$ as determined by \citet{Mosser2011} has a different definition, closer to the analytical asymptotic relation, and its value for RGB stars is systematically larger by $\sim 1\,\%$ compared to the one from individual mode frequencies, as shown on Fig. \ref{fig:Dnu_comp}. Meanwhile, there is no specific trend in the difference between the two $\deltanu$ estimates for RC stars.

Therefore, in Sect. \ref{sec:scaling_comp}, we use raw scaling relations in combination with $\deltanu$ from \citet{Mosser2011}; while, in Sects. \ref{sec:rel_comp} and \ref{sec:param_comp} where theoretically-motivated corrections to the $\deltanu$ scaling are used, we adopt $\deltanu$ from \citet{Yu2018} instead.

%__________________________________________________________________

\section{Detailed objectives}
\label{sec:objectives}

In order to simplify the statistical analysis of our results, we formulate the problem in the astrometric data space, i.e. parallax space. Significant biases can arise from the inversion of parallaxes into distances 
and from sample truncation, such as the removal of negative parallaxes and / or parallaxes with a relative error above a given threshold \citep{Luri2018}. Thus, in the current investigation, we avoid doing any of these. On the other hand, it is quite reasonable to invert asteroseismic distances to obtain parallaxes because their uncertainties are typically lower than a few percent \citep[see, e.g.,][]{Rodrigues2014}.

If one wishes to express the parallax as a function of the apparent and intrinsic luminosity of a star, this can be done using the Stefan-Boltzmann law, as follows: 
\begin{align}
	\varpi = c_{\lambda} \, \left(\frac{R_{\rm bb}}{R_{\rm \odot}} \right)^{-1} \left(\frac{T_{\rm eff}}{T_{\rm eff, \odot}} \right)^{-2} \, , \label{eq:plx_Gaia}
\end{align}
where $R_{\rm bb}$ is the radius of the black body of effective temperature $T_{\rm eff}$, i.e. the photospheric radius, and $c_{\lambda}=10^{-0.2 \, \left(m_{\lambda}+BC_{\lambda}+5-A_{\lambda}-M_{\rm bol,\odot}\right)}$. $m_{\lambda}$, $BC_{\lambda}$, and $A_{\lambda}$ are the magnitude, bolometric correction, and extinction in a given band $\lambda$, and we adopt $M_{\rm bol, \odot}=4.75$ for the Sun’s bolometric magnitude. Thereafter, we will resort to the 2MASS $K$-band magnitude properties ($m_{K}$, $BC_{K}$, and $A_{K}$), whenever we need to estimate the coefficient $c_{\lambda}$.

\subsection{Asteroseismic parallax}
\label{sec:parallax_seismo}

Engaging in such a parallax comparison requires a way to express the seismic information in terms of parallax, to be compared to the \textit{Gaia} astrometric measurements. Seismic parallaxes are also based on the Stefan-Boltzmann law. Going back to the foundations of ensemble asteroseismology, seismic scaling relations provide relevant estimates of the stellar masses and radii. 
From Eqs. (\ref{eq:Dnu}) and (\ref{eq:numax}), their expressions are as follows:
\begin{align}
	\left( \frac{M}{M_{\rm \odot}} \right) &\approx \left(\frac{\nu_{\rm max}}{\nu_{\rm max, \odot}} \right)^3 \left(\frac{\deltanu}{\deltanu_{\rm \odot}} \right)^{-4} \left(\frac{T_{\rm eff}}{T_{\rm eff,\odot}} \right)^{3/2} \, , \label{eq:mass_scaling} \\
	\left( \frac{R}{R_{\rm \odot}} \right) &\approx \left(\frac{\nu_{\rm max}}{\nu_{\rm max, \odot}} \right) \left(\frac{\deltanu}{\deltanu_{\rm \odot}} \right)^{-2} \left(\frac{T_{\rm eff}}{T_{\rm eff,\odot}} \right)^{1/2} \, , \label{eq:rad_scaling}
\end{align}
involving both global asteroseismic observables $\deltanu$ and $\numax$, and $T_{\rm eff}$. The solar references are taken as $\deltanu_{\rm \odot}=135 \ \mu$Hz, $\nu_{\rm max, \odot}=3090 \ \mu$Hz, and $T_{\rm eff,\odot}=5777 \ \rm K$. 
It is assumed here that the seismic radius, $R$, and the black-body radius, $R_{\rm bb}$, are the same. Finally, using Eqs. (\ref{eq:plx_Gaia}) and (\ref{eq:rad_scaling}), the seismic parallax ensuing from the scaling relations can be written as
\begin{align}
	\varpi_{\rm scaling} = c_{\lambda} \, \left(\frac{\nu_{\rm max}}{\nu_{\rm max, \odot}} \right)^{-1} \left(\frac{\deltanu}{\deltanu_{\rm \odot}} \right)^2 \left(\frac{T_{\rm eff}}{T_{\rm eff, \odot}} \right)^{-5/2} \, . \label{eq:plx_scaling}
\end{align}

The seismic scaling relations have been widely used, even though it is known that they are not precisely calibrated yet. Testing their validity has become a very active topic in asteroseismology, and has been addressed in several ways. 
It may take the form of a comparison between asteroseismic radii and independent measurements of stellar radii \citep[e.g.][]{Huber2012,Gaulme2016,Miglio2016,Huber2017}. 
An alternative approach consists in validating the relation between the average large frequency separation and the stellar mean density from model calculations \citep{Ulrich1986}. 
The asymptotic approximation for acoustic oscillation modes tells us that $\deltanu$ is directly related to the sound travel-time in the stellar interior, and therefore depends on the stellar structure \citep{Tassoul1980}. As mentioned in Sect. \ref{sec:intro}, Eq. (\ref{eq:Dnu}) is approximate and assumes that stars, in general, are homologous to the Sun and that the measured $\deltanu$ corresponds to $\deltanu$ in the asymptotic limit; in practice, that is not the case \citep[for further details see, e.g.,][]{Belkacem2013}. The sound speed in their interior (hence the total acoustic travel-time) does not simply scale with mass and radius only. In particular, whether a red-giant star is burning hydrogen in a shell or helium in a core, its internal temperature (hence sound speed) distribution will be different. This led several authors to quantify any deviation from the $\deltanu$ scaling relation these differences could cause \citep[e.g.][]{White2011,Miglio2012a,Belkacem2012,Guggenberger2016,Sharma2016,Rodrigues2017}. Stellar evolution calculations show that 
the deviation varies by a few percent with mass, chemical composition, and evolutionary state. That is why the seismic parallax can also be estimated from the large separation determined with grid-based modelling, i.e. statistical methods taking into account stellar theory predictions (e.g. allowed combinations of mass, radius, effective temperature, and metallicity) as well as other kinds of prior information (e.g. duration of evolutionary phases, star formation rate, initial mass function). In particular, the Bayesian tool \textsc{param} uses $\deltanu$, $\numax$, $T_{\rm eff}$, $\log g$, [Fe/H], [$\alpha$/Fe] (when available), and photometric measurements to derive probability density functions for fundamental stellar parameters, including distances \citep{Rodrigues2014,Rodrigues2017}. 

\begin{table*}[]
\centering
\begin{tabular}{ccccccc}
\toprule
Ev. state & $\varpi_{\rm Gaia}$ ($\mu$as) & $G$ & $\numax$ ($\mu$Hz) & $G_{\rm BP}-G_{\rm RP}$ & $M_{\rm PARAM}$ ($\mathrm{M_{\rm \odot}}$) & [Fe/H] \\
\midrule
RGB & 708 & 12.2 & 72.7 & 1.35 & 1.14 & $-$0.036  \\
RC & 625 & 11.8 & 40.0 & 1.29 & 1.34 & $-$0.0045 \\ 
\bottomrule
\end{tabular}
\caption{Values of $\overline{X}$ (mean value of the stellar parameter $X$) for RGB (top) and RC stars (bottom).}
\label{table:X}
\end{table*}

\begin{figure*}
   \centering
   \includegraphics[width=0.33\hsize]{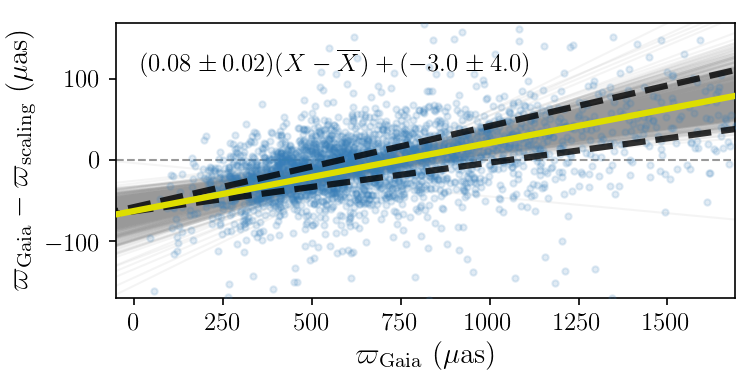}
   \includegraphics[width=0.33\hsize]{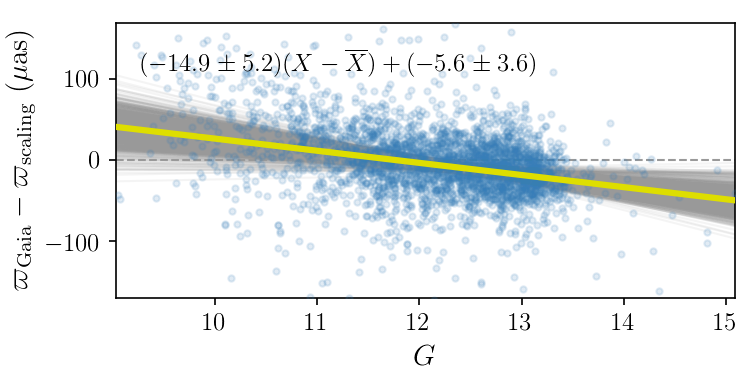}
   \includegraphics[width=0.33\hsize]{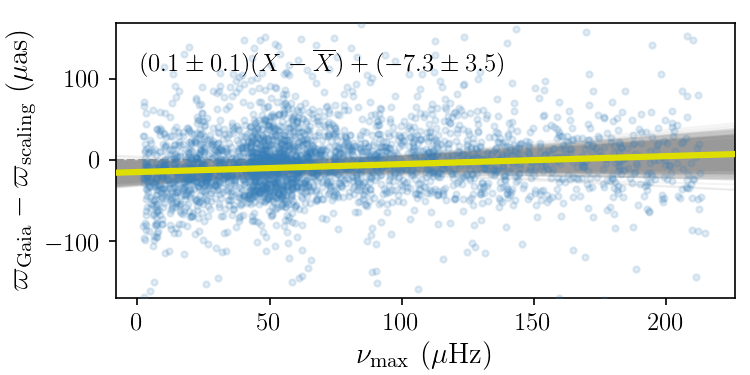}
   \includegraphics[width=0.33\hsize]{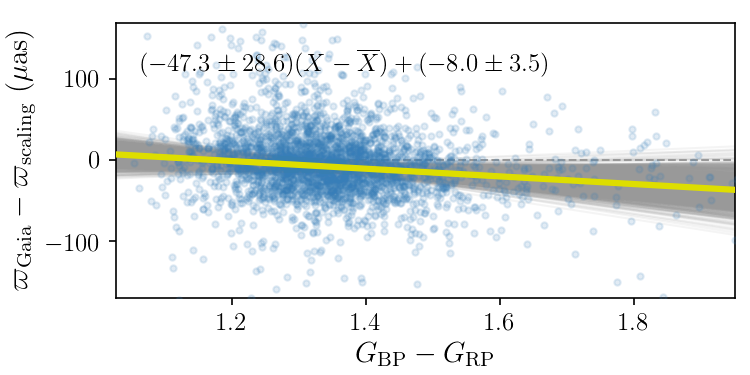}
   \includegraphics[width=0.33\hsize]{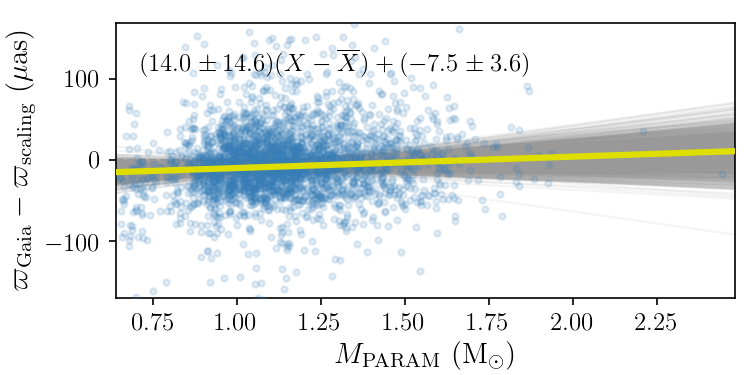}
   \includegraphics[width=0.33\hsize]{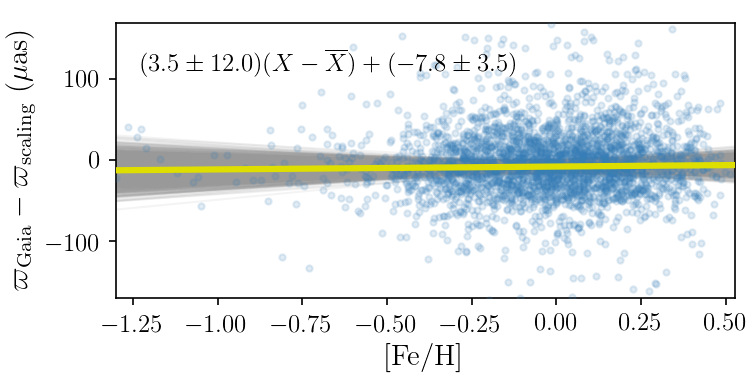}
   \caption{Parallax difference $\varpi_{\rm Gaia}-\varpi_{\rm seismo}$ for RGB stars, with the asteroseismic parallax derived from the raw scaling relations, as a function of $\varpi_{\rm Gaia}$, $G$, $\numax$, $G_{\rm BP}-G_{\rm RP}$, $M_{\rm PARAM}$, and [Fe/H]. The distribution of the $N$ realisations of the \textsc{ransac} algorithm is indicated by the grey-shaded region and the yellow line displays the average linear fit, for which the relation is given at the top of each subplot. The values of $\overline{X}$ for RGB stars are given in Table \ref{table:X}. The summary statistics are: $\left(\overline{\Delta \varpi}\right)_{\rm m} = -6.2 \pm 1.3 \ \mu$as, $\left(\overline{\Delta \varpi}\right)_{\rm w} = -7.9 \pm 0.8 \ \mu$as, and $z=0.89$. The black dashed lines correspond to the average linear fits when a $\pm 100 \ \rm K$ shift in $T_{\rm eff}$ is applied.}
   \label{fig:scaling_RGB}
\end{figure*}

\begin{figure*}
   \centering
   \includegraphics[width=0.33\hsize]{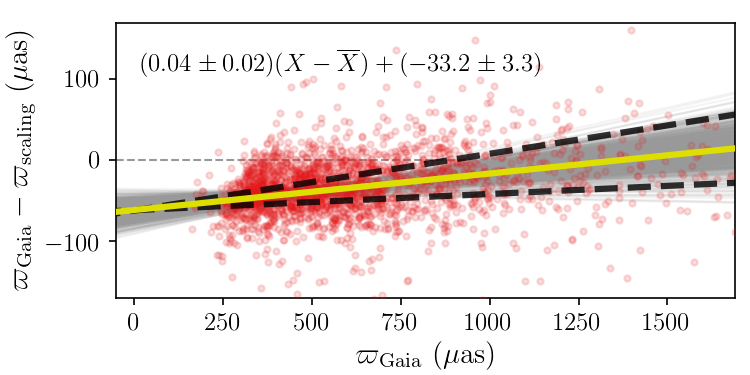}
   \includegraphics[width=0.33\hsize]{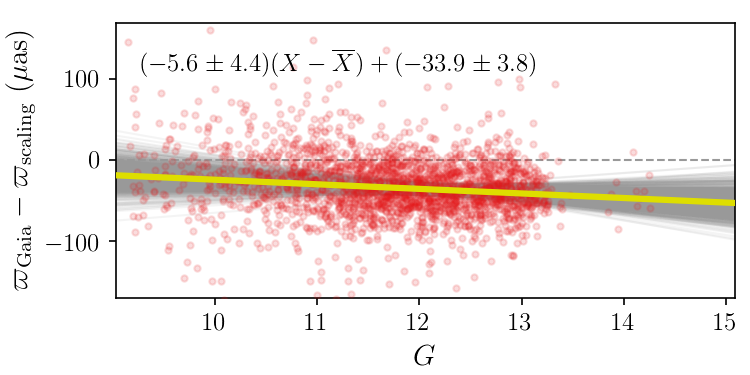}
   \includegraphics[width=0.33\hsize]{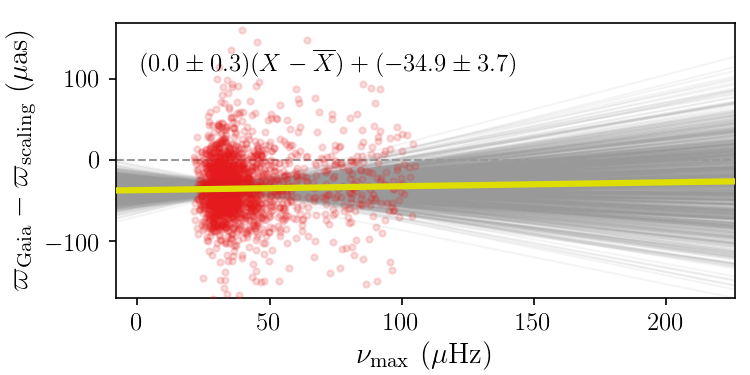}
   \includegraphics[width=0.33\hsize]{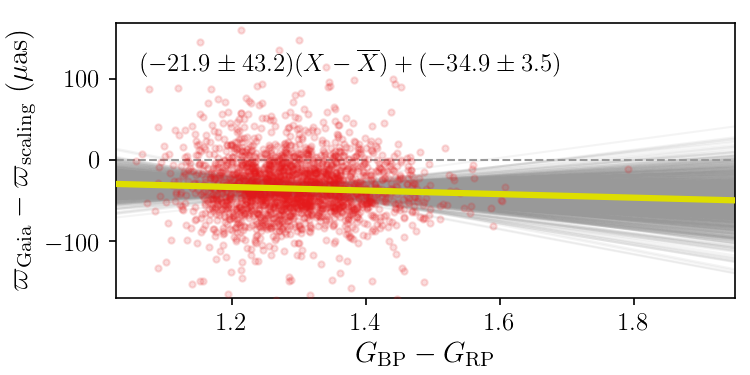}
   \includegraphics[width=0.33\hsize]{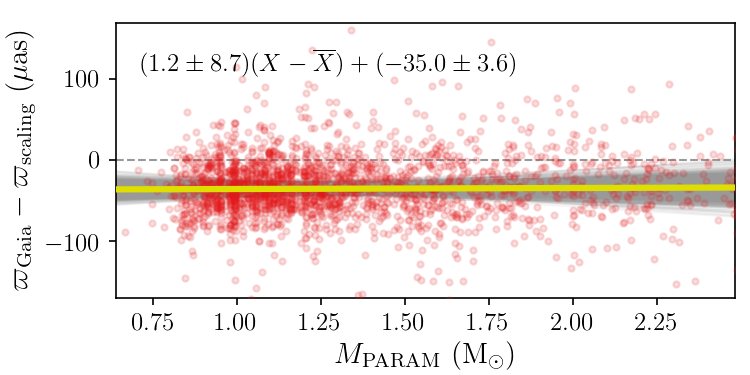}
   \includegraphics[width=0.33\hsize]{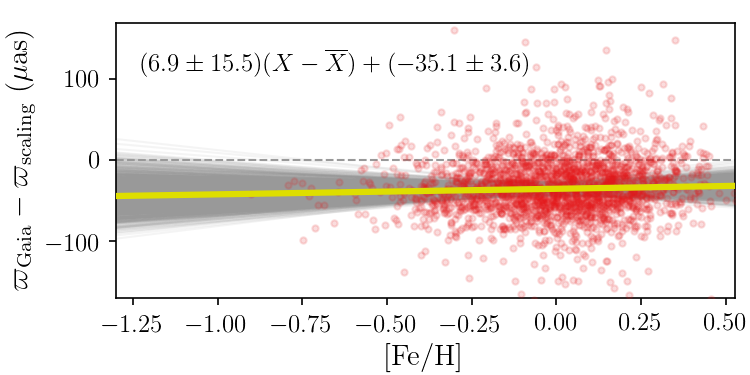}   
   \caption{Same as Fig. \ref{fig:scaling_RGB} for RC stars. The values of $\overline{X}$ for RC stars are given in Table \ref{table:X}. The summary statistics are: $\left(\overline{\Delta \varpi}\right)_{\rm m} = -34.6 \pm 1.4 \ \mu$as, $\left(\overline{\Delta \varpi}\right)_{\rm w} = -35.6 \pm 0.9 \ \mu$as, and $z=0.84$.}
   \label{fig:scaling_RC}
\end{figure*}

The asteroseismic results thus depend on the method employed. 
This aspect is explored in more detail in Sect. \ref{sec:kepler}, where three distinct seismic methods are tested (with the appropriate $\deltanu$, as discussed in Sect. \ref{sec:Dnu}): the raw scaling relations, a relative correction to the $\deltanu$ scaling between RGB and RC stars, and a model-grid-based Bayesian approach defining $\deltanu$ from individual frequencies. Furthermore, in Sect. \ref{sec:calibration}, we combine asteroseismic and astrometric data to simultaneously calibrate the scaling relations and the \textit{Gaia} zero-point. 

\subsection{Method}

Our study is based on the analysis of the absolute, rather than relative, difference between \textit{Gaia} and seismic parallaxes: $\Delta \varpi = \varpi_{\rm Gaia}-\varpi_{\rm seismo}$. This is for two reasons: first, the global zero-point offset in \textit{Gaia} parallaxes is absolute \citep{Lindegren2018}; second, working in terms of relative difference can 
amplify trends, e.g. due to offsets having a greater impact on small parallaxes. 

We explore the trends of the measured offset ($\Delta \varpi$) for a set of stellar parameters: the \textit{Gaia} parallax $\varpi_{\rm Gaia}$, the $G$-band magnitude, the frequency of maximum oscillation $\numax$, the $G_{\rm BP}-G_{\rm RP}$ colour index, the mass inferred from \textsc{param} ($M_{\rm PARAM}$), and the metallicity [Fe/H]. Each of these relations is described with a linear fit obtained through a \textsc{ransac} algorithm \citep{Fischler1981}. The fitting parameters' uncertainties are estimated by making $N=1000$ realisations of the set of parameters analysed with \textsc{ransac}, where a normally distributed noise is added using the observed uncertainties on $\Delta \varpi$ and the different stellar parameters. Because the fitting parameters are strongly dependent on the range of values covered by the independent variable $X$ (the stellar parameter considered), the fits are expressed in the following form: $\Delta \varpi (X) = \alpha_{X} (X-\overline{X}) + \beta_{X}$. $\alpha_{X}$ is the slope, $\beta_{X}$ is the intercept from which $\alpha_{X} \overline{X}$ was subtracted, and $\overline{X}$ is the mean value of the stellar parameter $X$ (Table \ref{table:X}). 

As part of the analysis, some summary statistics are also given: 
\begin{itemize}
    \item the median parallax difference $\left(\overline{\Delta \varpi}\right)_{\rm m}$ and the associated uncertainty $\sigma_{\left(\overline{\Delta \varpi}\right)_{\rm m}}$;
    \item the weighted average parallax difference
    \begin{align}
        \left(\overline{\Delta \varpi}\right)_{\rm w} = \frac{\sum_{i=1}^{N} \Delta \varpi_{\rm i} / \sigma_{\Delta \varpi_{\rm i}}^2}{\sum_{i=1}^{N} 1 / \sigma_{\Delta \varpi_{\rm i}}^2} \, ,
        \label{eq:wmean}
    \end{align}
    for which the uncertainty quoted is the weighted standard deviation, which gives a measure of the spread and also takes into account the individual (formal) uncertainties in $\Delta \varpi$, i.e.
    \begin{align}
        \sigma_{\left(\overline{\Delta \varpi}\right)_{\rm w}} = \sqrt{\frac{\sum_{i=1}^{N} \left(\Delta \varpi_{\rm i} - \left(\overline{\Delta \varpi}\right)_{\rm w}\right)^2 / \sigma_{\Delta \varpi_{\rm i}}^2}{(N-1) \sum_{i=1}^{N} 1 / \sigma_{\Delta \varpi_{\rm i}}^2}} \, ;
    \end{align}
    \item and the ratio $z=\sigma_{\left(\overline{\Delta \varpi}\right)_{\rm w}} \, / \, \overline{\sigma_{\left(\Delta \varpi\right)_{\rm w}}}$, where $\overline{\sigma_{\left(\Delta \varpi\right)_{\rm w}}}= 1 / \sqrt{\sum_{i=1}^{N} 1 / \sigma_{\Delta \varpi_{\rm i}}^2}$ is the uncertainty of the weighted mean estimated from the formal uncertainties on $\Delta \varpi$, which allows one to assess how well the formal fitting uncertainties reflect the scatter in the data; if the $\Delta \varpi$ scatter is dominated by random errors and the formal uncertainties reflect the true observational uncertainties, then $z$ is close to unity.
\end{itemize}

In the following, unless stated otherwise, the weighted average parallax difference estimator will be used for the offsets quoted in the text.

%__________________________________________________________________

\section{Analysis of the \textit{Kepler} field}
\label{sec:kepler}

In this section, we focus on the comparison between \textit{Gaia} DR2 and asteroseismology for red giants in the \textit{Kepler} field. We take advantage of \citet{Elsworth2017}'s classification method, based on the structure of the dipole-mode oscillations of mixed character, to distinguish between shell-hydrogen burning stars, on the red-giant branch, and core-helium burning stars, in the red clump (including secondary clump stars as well). From the asteroseismic perspective, three different approaches are employed in order to emphasise the influence of the seismic method on the measured parallax zero-point.

\subsection{Raw scaling relations}
\label{sec:scaling_comp}

\begin{figure*}
   \centering
   \includegraphics[width=0.8\hsize]{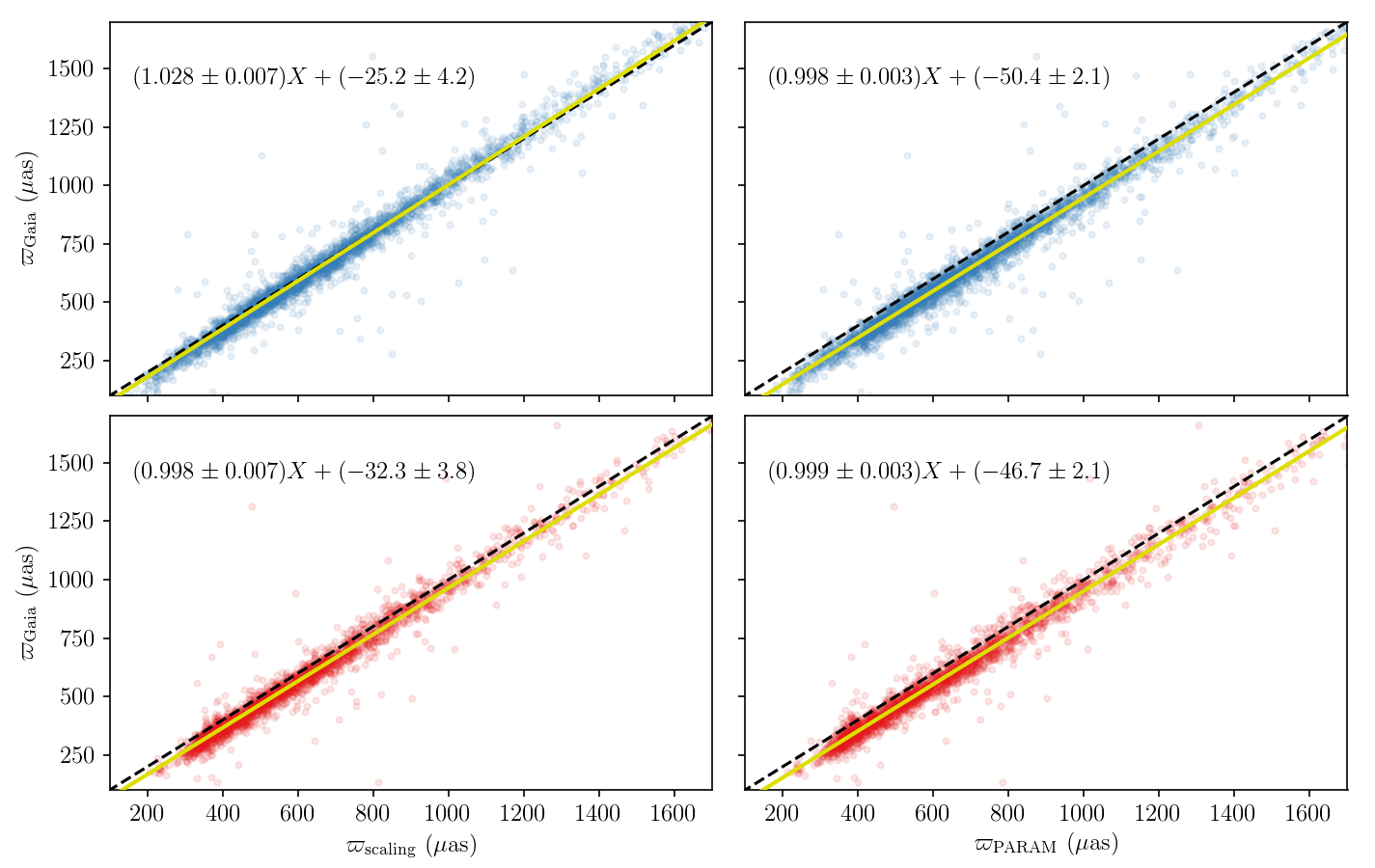}
   \caption{$\varpi_{\rm Gaia}$ as a function of $\varpi_{\rm scaling}$ (left) and $\varpi_{\rm PARAM}$ (right), for RGB (top) and RC (bottom) stars. The yellow line displays the linear fit, averaged over $N$ realisations, for which the relation is given at the top of each subplot. The black dashed line indicates the 1:1 relation.}
   \label{fig:scaling_gaia}
\end{figure*}

We start with the raw scaling relations, to which no correction has been applied: the seismic parallax is directly estimated from Eq. (\ref{eq:plx_scaling}), using $\deltanu$ from \citet{Mosser2011}. The comparison with \textit{Gaia} parallaxes is shown on Figs. \ref{fig:scaling_RGB} and \ref{fig:scaling_RC} for RGB and RC stars, respectively. At first sight, we observe a strong dependence of the RGB parallax difference with $\varpi_{\rm Gaia}$: as the latter increases, $\Delta \varpi$ significantly increases as well (Fig. \ref{fig:scaling_RGB}, top left panel). Such a trend also appears for RC stars, but to a much lesser extent. In fact, the parallax zero-point is expected to show variations depending on position, magnitude and colour \citep{Lindegren2018} but not on parallax, unless we consider stars with the same intrinsic luminosity (e.g. RC stars; see Fig. \ref{fig:scaling_RC}). In such a case, the dependence on apparent magnitude would manifest itself as a dependence on parallax. In addition, for core-helium burning stars, one can see that there is a cut-off at low astrometric parallaxes ($\varpi_{\rm Gaia}<350 \ \mu$as; Fig. \ref{fig:scaling_RC}, top left panel). This is most likely related to the selection in magnitude in \textit{Kepler} \citep[see, e.g.,][]{Farmer2013}, which translates into a limit on distance. \textit{Gaia} parallaxes, having larger uncertainties compared to their asteroseismic counterparts, can lead to distances greater than this limit and, when represented on the $x$-axis, create a horizontal structure (adding to the vertical structure caused by the scatter of the parallax difference) as observed. On the contrary, if we had the seismic parallax on the $x$-axis instead, such a structure would disappear and the slope would become flatter. Still, we note that these trends might either come from the seismic parallax or from the correlation between the parallax difference and the \textit{Gaia} parallax. Having a deeper look at the summary statistics, there seems to be a considerable difference in the measured offset: that of RGB stars reaches up to $-8 \ \mu$as, compared to RC stars displaying an average value of $-36 \ \mu$as. On their own, these results could be interpreted as a minimal difference between astrometric and asteroseismic measurements for stars along the RGB, but there remains the issue of the apparent trends of $\Delta \varpi$ with parallax.

\begin{figure}
   \centering
   \includegraphics[width=0.8\hsize]{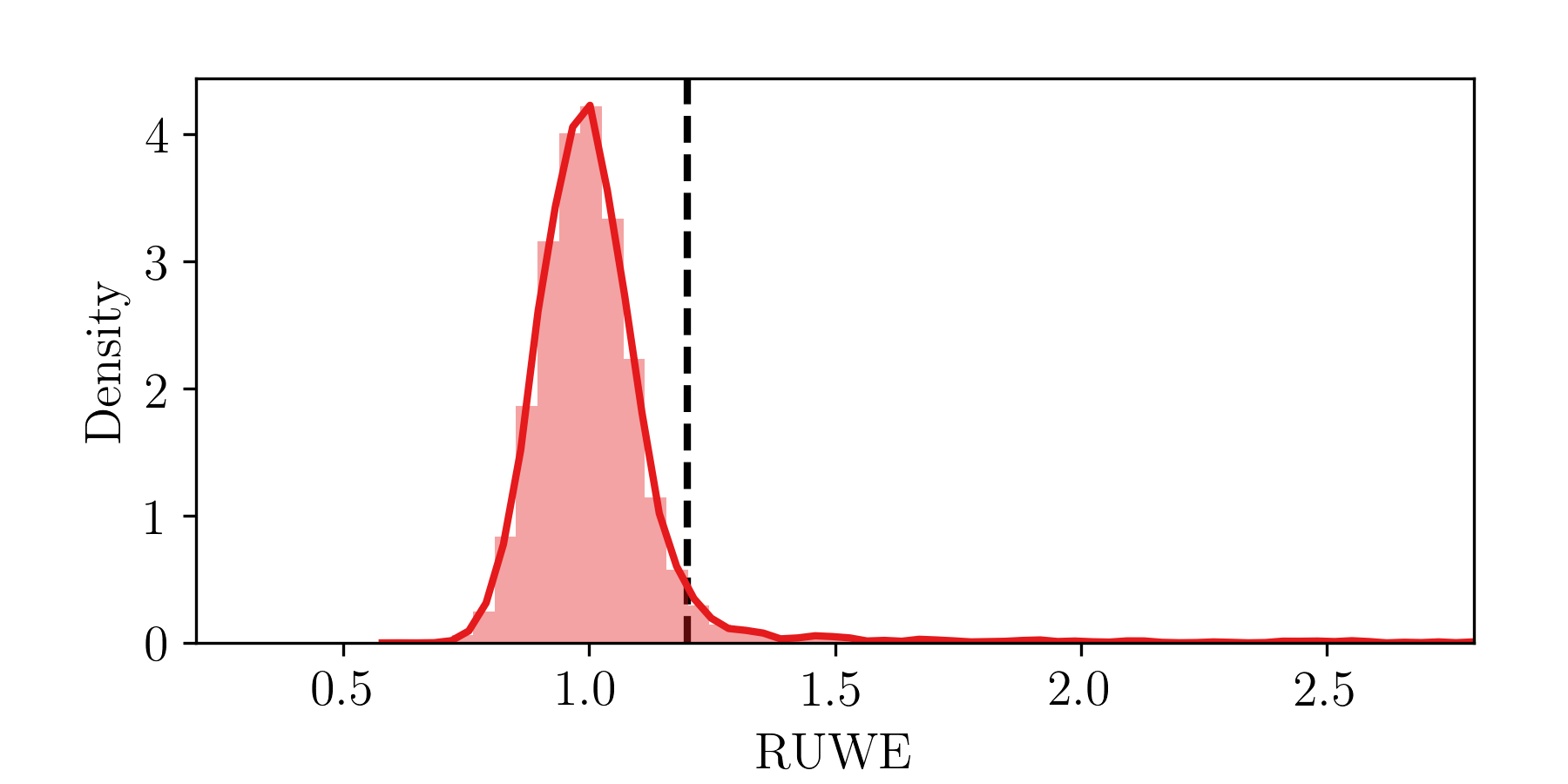}
   \caption{Renormalised Unit Weight Error distribution for \textit{Kepler} RGB and RC stars. The vertical dashed line indicates the threshold adopted: $\rm{RUWE}\leq 1.2$.}
   \label{fig:kepler_RUWE}
\end{figure}

To clarify this situation, we show on Fig. \ref{fig:scaling_gaia} the relation between the seismic and \textit{Gaia} parallaxes, separately for RGB and RC stars. While the latter display a relation nearly parallel to the 1:1 line, 
RGB stars show a slope of $1.028 \pm 0.007$, which is significantly different from 1 and is not solely due to a correlation effect. Because all sources are treated as single stars in \textit{Gaia} DR2, the results for resolved binaries may sometimes be spurious due to confusion of the components \citep{Lindegren2018}. The Renormalised Unit Weight Error (RUWE) 
is recommended as a goodness-of-fit indicator for \textit{Gaia} DR2 astrometry (see the technical note \textit{GAIA-C3-TN-LU-LL-124-01} available from the DPAC Public Documents page\footnote{\url{https://www.cosmos.esa.int/web/gaia/public-dpac-documents}}). 
It is computed from the following quantities:
\begin{itemize}
    \item $\chi^2=$ \texttt{astrometric\_chi2\_al};
    \item $N=$ \texttt{astrometric\_n\_good\_obs\_al};
    \item $G$;
    \item and $G_{\rm BP}-G_{\rm RP}$.
\end{itemize}
Figure \ref{fig:kepler_RUWE} shows the distribution of the RUWE for stars in the \textit{Kepler} field, including both RGB and RC evolutionary phases (their distinction does not affect the shape of the distribution). Because there seems to be a breakpoint around $\rm{RUWE}=1.2$ between the expected distribution for well-behaved solutions and the long tail towards higher values, we adopt $\rm{RUWE}\leq 1.2$ as a criterion for ``acceptable'' solutions. By imposing this condition, the scatter is reduced, but the slope appearing in Fig. \ref{fig:scaling_gaia} is still present and the offset remains unchanged. Therefore, this steep slope could potentially be a symptom of biases in the seismic scaling relations. To a good approximation, the changes in the slope due to changes in $T_{\rm eff}$, or modifications in the scaling relations, can be obtained as linear perturbations of Eq. (\ref{eq:plx_scaling}). The question of their calibration using \textit{Gaia} data is addressed in Sect. \ref{sec:calibration}. In addition, the similar distributions of the RUWE for RGB and RC stars also point in favour of the fact that the quality of \textit{Gaia} parallaxes is not responsible for the different behaviour of these stars in Figs. \ref{fig:scaling_RGB} and \ref{fig:scaling_RC}.

A fairly strong trend also appears for the parallax difference as a function of the $G$-band magnitude, especially in the case of RGB stars (Fig. \ref{fig:scaling_RGB}, top middle panel). Within \textit{Gaia} itself, observations are acquired in different instrumental configurations and need to be calibrated separately depending on, e.g., the window class and the gate activation \citep[for further details see, e.g.,][]{Riello2018}. In the range of magnitudes covered by red-giant stars, several changes occur:
\begin{itemize}
    \item at $G=11.5$, the BP/RP (Blue Photometer/Red Photometer) window class switches from 2D to 1D;
    \item at $G=12$, there is the transition between gated (to avoid saturation affecting bright sources) and ungated observations;
    \item at $G=13$, the AF (Astrometric Field) window class changes from 2D to 1D.
\end{itemize}
In order to separate the different effects, we divide the RGB and RC samples in bins of $G$ as follows: $G\leq11.5$, $G\in\,[11.5,12]$, $G\in\,[12,13]$, and $G>13$. Doing so results in an offset decreasing from $16$ to $-21 \ \mu$as between the lowest and the highest bins for RGB stars, and from $-28$ to $-44 \ \mu$as for RC stars. 
We remark that the trend of the offset with increasing magnitude is negative in both cases, but stronger for RGB stars than for RC stars (see top middle panels in Figs. \ref{fig:scaling_RGB} and \ref{fig:scaling_RC}), which might again indicate a problem regarding the raw scaling relations. $\Delta \varpi$ does not seem to exhibit any noteworthy relation with the other stellar parameters. Hence, they will not be shown throughout the rest of the paper, apart from $\numax$ which is an important asteroseismic indicator of the evolutionary stages.

\begin{figure}
   \centering
   \includegraphics[width=0.99\hsize]{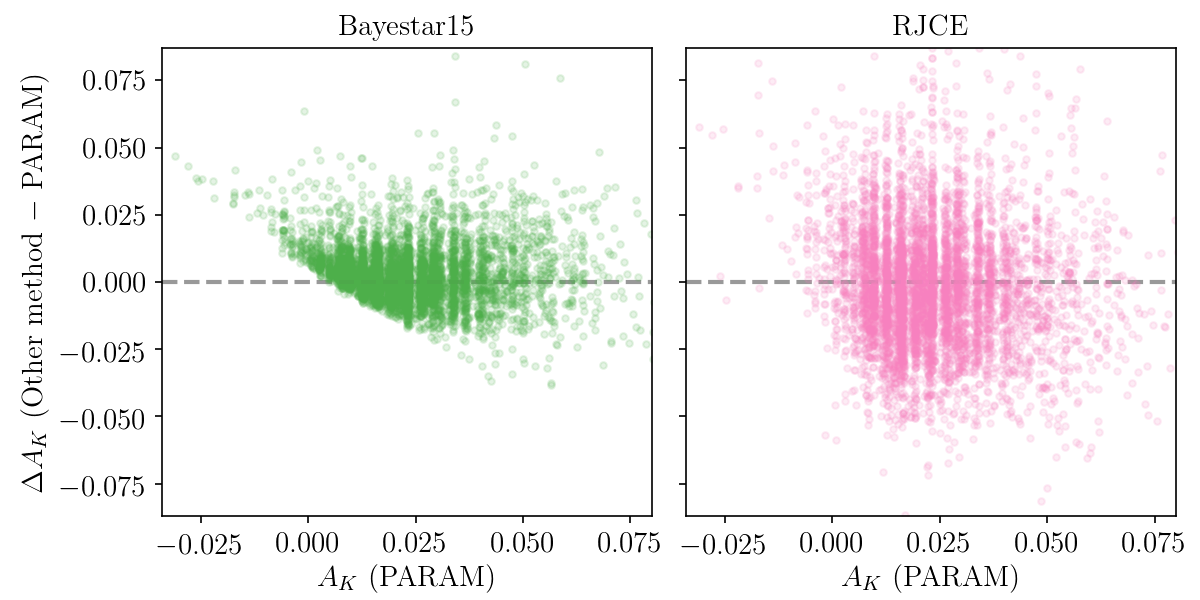}
   \caption{Absolute difference in $A_K$ between the \citet{Green2015} map (left) / RJCE method (right) and \textsc{param}, as a function of \textsc{param}'s extinctions.}
   \label{fig:ak_comp}
\end{figure}

\begin{figure*}
   \centering
   \includegraphics[width=0.33\hsize]{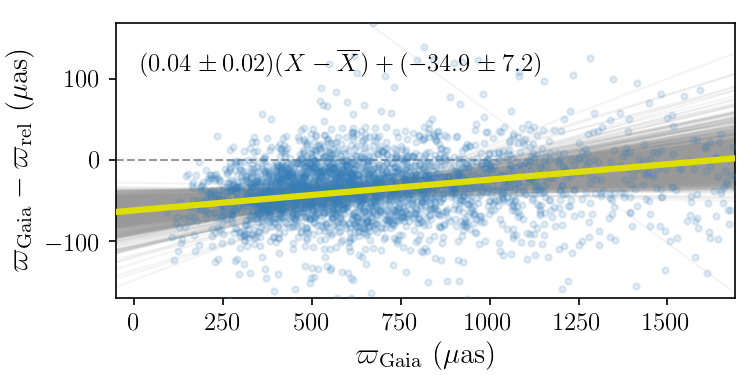}
   \includegraphics[width=0.33\hsize]{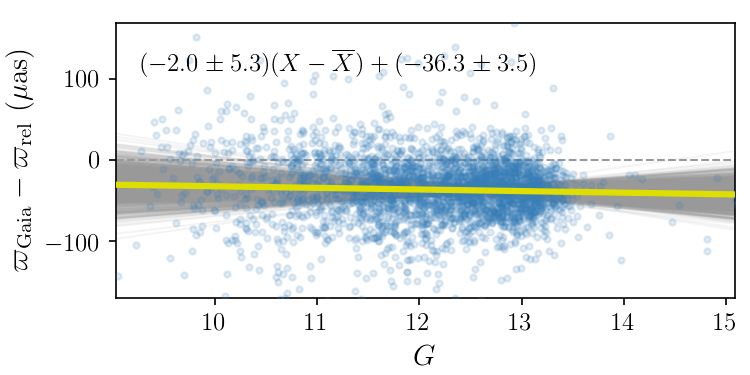}
   \includegraphics[width=0.33\hsize]{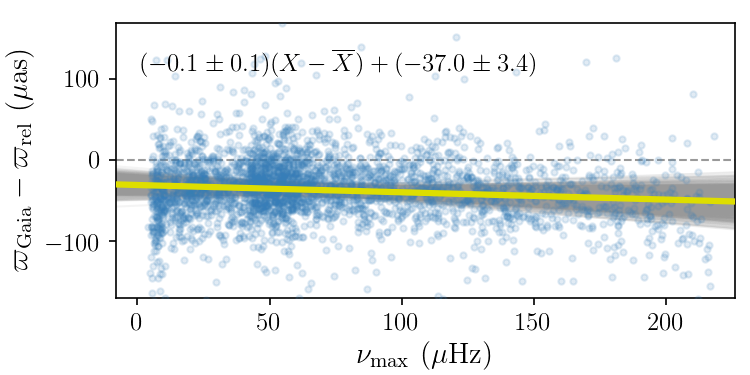}
   \caption{Same as Fig. \ref{fig:scaling_RGB} with the asteroseismic parallax derived from the scaling relations, where a 2.7\% correction factor has been applied to the $\deltanu$ scaling. The summary statistics are: $\left(\overline{\Delta \varpi}\right)_{\rm m} = -35.8 \pm 1.3 \ \mu$as, $\left(\overline{\Delta \varpi}\right)_{\rm w} = -35.5 \pm 0.8 \ \mu$as, and $z=0.88$.}
   \label{fig:emp_RGB}
\end{figure*}

Despite the above, a considerable advantage of using the raw scaling relations is to give us the agility and flexibility to have a direct test of potential systematic effects. Besides, later we will show that a lot of the departures of the slopes from unity can be removed by using grid-based modelling (Sect. \ref{sec:param_comp}). Here, we explore the influence that other non-seismic inputs, i.e. the effective temperature scale and the extinctions, may have on the comparison. 
$T_{\rm eff}$ appears explicitly in Eq. (\ref{eq:plx_scaling}), but also implicitly through the bolometric correction contained in the coefficient $c_{\lambda}$, the definition of which is given in Sect. \ref{sec:objectives}. The combination of these two factors leads to an increase (decrease) of both the $\varpi_{\rm scaling}$---$\varpi_{\rm Gaia}$ slope coefficient and the offset with increasing (decreasing) $T_{\rm eff}$: a $\pm 100 \ \rm K$ shift results in a $\pm$10-15 $\mu$as variation in the parallax difference. Reducing the effective temperature by $100$ K is almost enough to obtain a slope of $\sim 1$, but not to have an offset in agreement with the red clump. 
On the other hand, setting the extinctions to zero or doubling their values barely affects the parallax difference, at the order of $\pm 6 \ \mu$as at the most. As a check, in Fig. \ref{fig:ak_comp}, we compare our extinction values with those derived by \citet{Green2015} (Bayestar15) and from the RJCE method \citep{Majewski2011}, to see if they are consistent with each other. For the most part, the differences are within the $\pm 0.02$ level, with a larger scatter on the RJCE's side. The typical (median) uncertainties on the extinctions are 0.007, 0.002, and 0.025 for \textsc{param}, Bayestar15, and RJCE, respectively. 
At low extinction values ($A_K < 0.025$), Bayestar15's extinctions are systematically larger, introducing a diagonal shape in the distribution: this is most likely a truncation effect caused by the fact that Bayestar15 only provides strictly non-negative extinction estimates, while \textsc{param} derives both positive and negative $A_K$ \citep{Rodrigues2014}. Such differences are not expected to significantly affect our comparison, as already implied by the above tests: the parallax offsets measured with these extinctions only differ by approximately $\pm 2 \ \mu$as. This is also partly due to the fact that we are using an infrared passband, which reduces the impact of reddening.

\subsection{Corrected $\deltanu$ scaling relation}
\label{sec:rel_comp}

From theoretical models, one expects that deviations from the $\deltanu$ scaling relation depend on mass, chemical composition, and evolutionary state, as discussed in Sect. \ref{sec:parallax_seismo}. Nevertheless, at fixed mass and metallicity (e.g. for a cluster), one can derive a relative correction to the scaling between RGB and RC stars, modifying Eq. (\ref{eq:Dnu}) as follows: $\deltanu' = \caldnu \deltanu$, where $\caldnu$ is a correction factor. This was done for the open cluster NGC 6791: \citet{Miglio2012a} compared asteroseismic and photometric radii, while \citet{Sharma2016} estimated $\caldnu$ along each stellar track of a grid of models \citep[see also][]{Rodrigues2017}. Both found a relative correction of $\sim 2.7 \, \%$ between the two evolutionary stages, $\caldnu$ being larger and closer to unity for RC stars. The value of $2.7 \, \%$ corresponds to the low-mass end ($M \sim 1.1 \ \rm M_{\rm \odot}$); the relative correction would be of the order of $2.5 \, \%$ for $1.2$-$1.3 \ \rm M_{\rm \odot}$ stars. 

Hence, as a first-order approximation, we apply this correction to our RGB sample: namely, for each star, we reduce $\deltanu$ from the scaling relation by 2.7\,\%. 
As this correction is based on a definition of $\deltanu$ from individual frequencies, it makes sense to use $\deltanu$ from \citet{Yu2018} to ensure consistency. 
Fig. \ref{fig:emp_RGB} shows how the inclusion of this correction from modelling affects the comparison for stars along the RGB. After applying the relative correction, the estimated offset becomes $-35 \ \mu$as, which is much closer to what was obtained with RC stars (Fig. \ref{fig:scaling_RC}), possibly indicating the relevance of the correction. 
However, even if the relations of $\Delta \varpi$ with the \textit{Gaia} parallax and the $G$-band magnitude seem flatter, the $\varpi_{\rm rel}$---$\varpi_{\rm Gaia}$ relation now displays a slope of $0.984 \pm 0.007$. This is most likely due to the application of an average correction initially derived for NGC 6791, and not quite suitable for the wide range of masses and metallicities covered by the sample. Finally, to help quantifying the effect of this correction, we also estimate the slope and the parallax offset using $\deltanu$ from \citet{Mosser2011}, as in the previous section, and we obtain $0.974$ and $-43 \ \mu$as. Thus, the dominant effect here is that of the correction, rather than the change of $\deltanu$.

\begin{figure*}
   \centering
   \includegraphics[width=0.33\hsize]{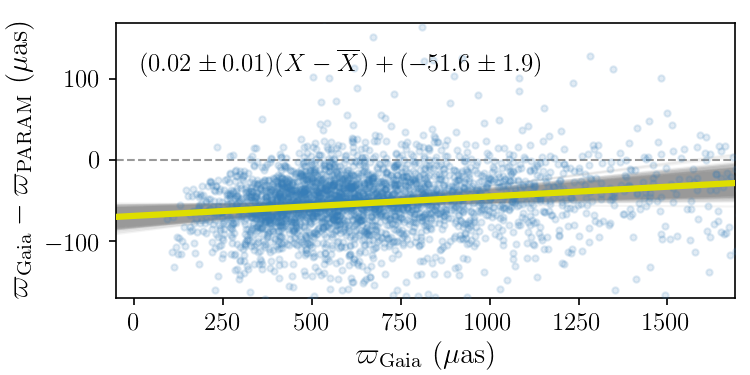}
   \includegraphics[width=0.33\hsize]{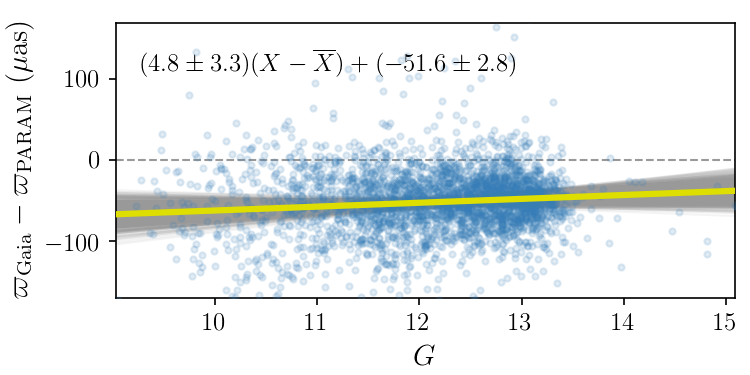}
   \includegraphics[width=0.33\hsize]{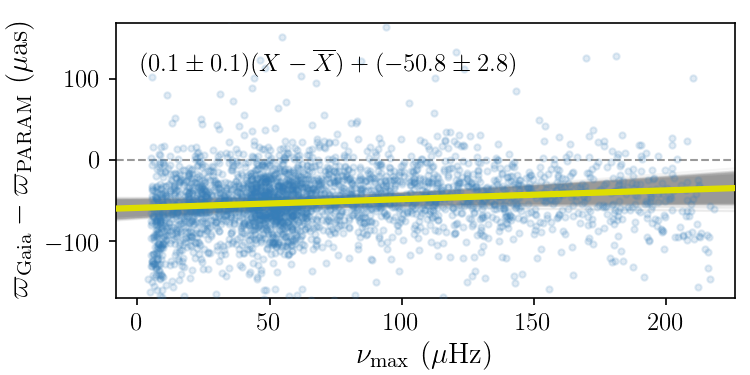}
   \caption{Same as Fig. \ref{fig:scaling_RGB} with the asteroseismic parallax derived from \textsc{param} \citep{Rodrigues2017}. The summary statistics are: $\left(\overline{\Delta \varpi}\right)_{\rm m} = -51.4 \pm 1.0 \ \mu$as, $\left(\overline{\Delta \varpi}\right)_{\rm w} = -51.7 \pm 0.8 \ \mu$as, and $z=1.24$.}
   \label{fig:param_RGB}
\end{figure*}

\begin{figure*}
   \centering
   \includegraphics[width=0.33\hsize]{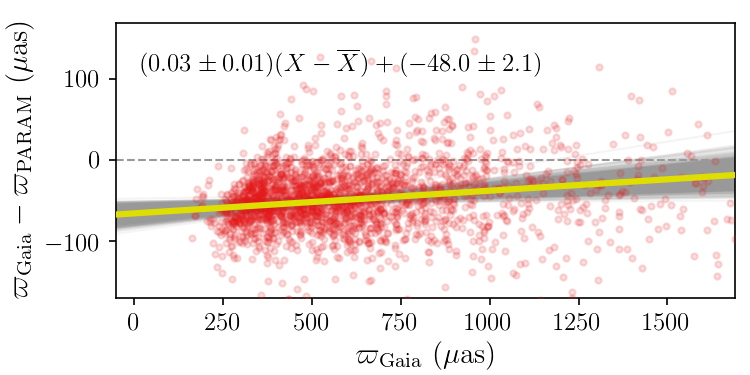}
   \includegraphics[width=0.33\hsize]{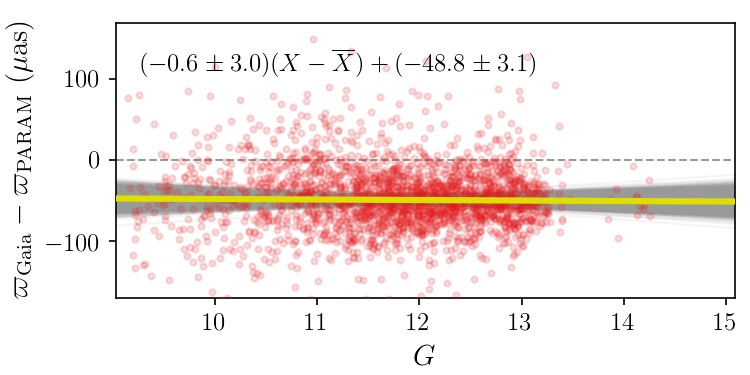}
   \includegraphics[width=0.33\hsize]{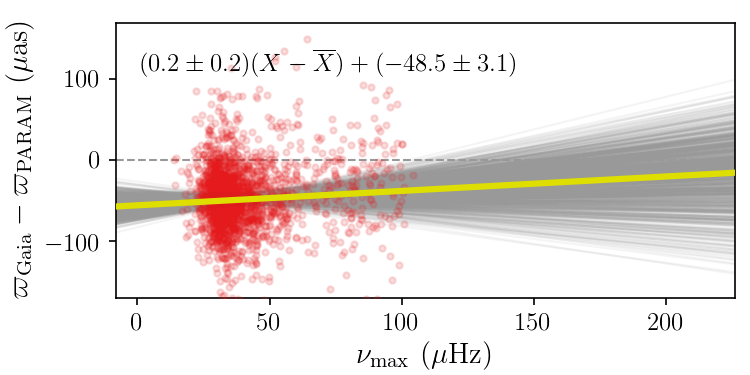}
   \caption{Same as Fig. \ref{fig:param_RGB} for RC stars. The summary statistics are: $\left(\overline{\Delta \varpi}\right)_{\rm m} = -48.3 \pm 1.1 \ \mu$as, $\left(\overline{\Delta \varpi}\right)_{\rm w} = -47.9 \pm 0.9 \ \mu$as, and $z=1.23$.}
   \label{fig:param_RC}
\end{figure*}

\subsection{$\deltanu$ from individual frequencies: grid-modelling}
\label{sec:param_comp}

With their Bayesian tool \textsc{param}, \citet{Rodrigues2017} replaced the $\deltanu$ scaling relation with an average large frequency definition stemming from a linear fitting of the individual radial-mode frequencies computed along the evolutionary tracks of the grid. Similarly, and as stated above, we use $\deltanu$ as estimated by \citet{Yu2018} for consistency with the $\deltanu$ definition adopted in the models in \textsc{param}. At this time, this approach has yielded masses / radii that show no systematic deviations to within a few percent of independent estimates (see, e.g., \citealt{Miglio2016}; \citealt{Handberg2017}; \citealt{Rodrigues2017}; \citealt{Brogaard2018}, who partially revisited the work by \citealt{Gaulme2016}). 
This method requires the use of a grid of models covering the complete relevant range of masses, ages, and metallicities. It is worth 
emphasising that the physical inputs of the models play a crucial role in the determination of stellar parameters via a Bayesian grid-based method. There is no absolute set of stellar models, and a few changes in their ingredients may also affect the outcome of an investigation such as ours. For details about the models considered here, we refer the reader to \citet{Rodrigues2017}, with the exception that we include element diffusion.

The comparison of the \textit{Gaia} parallaxes with the seismic ones estimated with \textsc{param} appears on Figs. \ref{fig:param_RGB}, for RGB stars, and \ref{fig:param_RC}, for RC stars. 
Both evolutionary phases have a flattened relation with $\varpi_{\rm Gaia}$, such that they now display similar slopes. In the RGB sample, the $\varpi_{\rm PARAM}$---$\varpi_{\rm Gaia}$ relation has a slope nearly equal to unity: $0.998 \pm 0.003$; that of RC stars is largely unchanged (see Fig. \ref{fig:scaling_gaia}). These effects bring the parallax zero-points really close: $-52$ and $-48 \ \mu$as for RGB and RC stars, respectively. The trends with $G$ are also relatively flat, resulting in small fluctuations as we move from low to high $G$ magnitudes: from $-58$ to $-51 \ \mu$as for stars on the RGB, and from $-46$ to $-52 \ \mu$as in the clump. These findings are reassuring in the sense that, if we were to find a trend with parallax or an evolutionary-state dependent offset, the issue would be down to seismology. Since we do not observe such effects, it seems relevant to use \textsc{param} with appropriate constraints to derive asteroseismic parallaxes. If we were to combine \textsc{param} with $\deltanu$ estimates from \citet{Mosser2011} instead, the RGB and RC offsets would become $-62$ and $-46 \ \mu$as, respectively. This introduces a significant RGB / RC relative difference in the parallax zero-point, which is neither due to the presence of secondary clump stars, nor to the different $\numax$ ranges covered by RGB and RC stars. Furthermore, the effect on the slopes is a decrease for RGB stars (0.986) and a very mild increase for RC stars (1.002). Again, these findings highlight the importance of ensuring consistency in the $\deltanu$ definition between observations and models. 

What follows below aims at quantifying how sensitive the findings with \textsc{param} are on additional systematic biases such as changes in the $T_{\rm eff}$ and [Fe/H] scales, and the use of different model grids. We tested that a $\pm 100 \ \rm K$ shift in $T_{\rm eff}$ affects $\Delta \varpi$ by $\pm 3 \ \mu$as for RGB stars, but this left the results largely unchanged for RC stars. It is not surprising that the order of magnitude of these variations  is lower compared to when we used the scaling relations at face value ($\pm 10$-$15 \ \mu$as). The grid of models restricts the possible range of $T_{\rm eff}$ values for a star with a given mass and metallicity, even more so when dealing with the very localised core-helium burning stars. Then, a $\pm 0.1 \ \rm dex$ shift in [Fe/H] affects $\Delta \varpi$ by $\mp 4$ and $\mp 2 \ \mu$as for RGB and RC stars, respectively. 
Finally, when considering models computed without diffusion \citep[described in][]{Rodrigues2017}, the parallax zero-point slightly increases for RGB stars ($\Delta \varpi \sim -57 \ \mu$as) and decreases for RC stars ($\Delta \varpi \sim -41 \ \mu$as), enhancing the discrepancy between the two evolutionary phases. Grids of models computed with and without diffusion differ, e.g., in terms of the mixing-length parameter, and of the initial helium and heavy-elements mass fractions obtained during the calibration of a solar model. Because of these combined effects, it is complex to interpret the respective impacts on RGB and RC stars. Since models with diffusion are in better agreement with, e.g., the helium abundance estimated in the open cluster NGC 6791 \citep{Brogaard2012} and constraints on the Sun from helioseismology \citep{Christensen-Dalsgaard2002}, they are our preferred choice for the current study. However, we stress that, at this level of precision, uncertainties related to stellar models are non-negligible (see Miglio et al. in prep.). 

\begin{figure}
   \centering
   \includegraphics[width=\hsize]{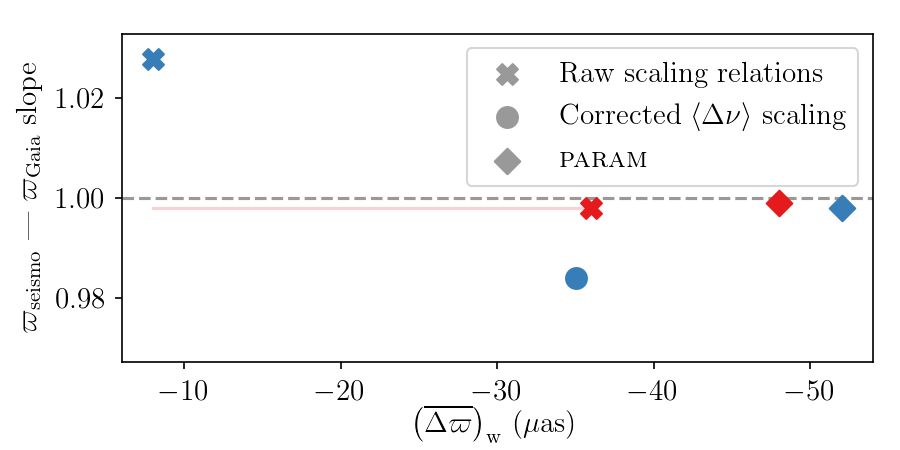}
   \caption{Slope of the various $\varpi_{\rm seismo}$---$\varpi_{\rm Gaia}$ relations --- using raw scaling relations (crosses; Sect. \ref{sec:scaling_comp}), a corrected $\deltanu$ scaling relation (circles; Sect. \ref{sec:rel_comp}), and \textsc{param} (diamonds; Sect. \ref{sec:param_comp}) --- as a function of the weighted average parallax difference for RGB (blue) and RC (red) stars. For RC stars, the line is extended to lower offset values for a better visualisation of how the slope compares with RGB stars when using scaling relations at face value.}
   \label{fig:summary}
\end{figure}

As a summary, Fig. \ref{fig:summary} shows how the $\varpi_{\rm seismo}$---$\varpi_{\rm Gaia}$ slopes and parallax offsets (weighted average parallax difference) evolve as we move from the scaling relations taken at face value to grid-based modelling, illustrating the convergence of RGB and RC stars both in terms of slope and offset when using \textsc{param}.

\subsection{External validation with open clusters}

We perform an external validation of our findings by using independent measurements for the open clusters NGC 6791 and NGC 6819, both in the \textit{Kepler} field. We adopt the distances given by eclipsing binaries: $d_{\rm NGC \ 6791}=4.01 \pm 0.14$ kpc \citep{Brogaard2011} and $d_{\rm NGC \ 6819}=2.52 \pm 0.15$ kpc \citep{Handberg2017}. The comparison with \textit{Gaia} DR2 parallax measurements \citep{Cantat-Gaudin2018} gives offsets of $-60.6 \pm 8.9 \ \mu$as and $-40.4 \pm 23.6 \ \mu$as for the former and the latter, respectively. This is reassuring as it is in line with the results obtained with \textsc{param}. 

\subsection{Influence of spatial covariances}
\label{sec:covariances}

As discussed by \citet{Lindegren2018} \citep[see also][]{Arenou2018}, spatial correlations are present in the astrometry, leading to small-scale systematic errors. The latter have a size comparable to that of the focal plane of \textit{Gaia}, i.e. $\sim 0.7^\circ$. In comparison, the \textit{Kepler} field, with an approximate radius of 7$^\circ$, is very large. The uncertainty on the inferred parallax offset may be largely underestimated, unless one takes these spatial correlations into account. For this reason, we perform a few tests in order to quantify how the various quantities derived in this work (average parallax difference and slope of the linear fits) and their uncertainties would be affected by the presence of spatial covariances. To be as representative as possible, in terms of spatial and distance distributions, we choose to work with our \textit{Kepler} sample. 

We first consider the seismic parallaxes estimated with \textsc{param} as the ``true'' parallaxes ($\varpi_{\rm true}$). This is an arbitrary choice and, thereafter, $\varpi_{\rm true}$ has to be viewed as a synthetic set of true parallaxes, completely independent from seismology. From there, we have to compute synthetic seismic and astrometric parallaxes. The former are calculated using the observed uncertainties on \textsc{param} parallaxes ($\sigma_{\varpi_{\rm seismo}}$):
\begin{align}
    \varpi_{\rm seismo} = \varpi_{\rm true} + \mathcal{N}(0, \sigma_{\varpi_{\rm seismo}}^2) \, ,
    \label{eq:sim_seismo}
\end{align}
where $\mathcal{N}(0, \sigma_{\varpi_{\rm seismo}}^2)$ is a normal distribution with mean zero and variance $\sigma_{\varpi_{\rm seismo}}^2$. Then, two sets of astrometric parallaxes are simulated, using the observed uncertainties on \textit{Gaia} parallaxes ($\sigma_{\varpi_{\rm Gaia}}$):
\begin{align}
    \varpi_{\rm Gaia}^{\rm unc} &= \varpi_{\rm true} + \mathcal{N}(0, \sigma_{\rm \varpi_{\rm Gaia}}^2) + \ogaia \, , \label{eq:sim_unc} \\
    \varpi_{\rm Gaia}^{\rm cor} &= \varpi_{\rm true} + \mathcal{N}(0, \sigma_{\rm \varpi_{\rm Gaia}}^2) + \mathcal{N}(\ogaia, \mathcal{S}) \, , \label{eq:sim_cor}
\end{align}
where $\ogaia =-50 \ \mu$as represents the \textit{Gaia} parallax zero-point and is set arbitrarily following our findings (Sect. \ref{sec:param_comp}), and $\mathcal{N}(\ogaia, \mathcal{S})$ is a multivariate normal distribution with mean $\ogaia$ and covariance matrix $\mathcal{S}$. These two parallaxes contain the same random error component ($\mathcal{N}(0, \sigma_{\rm \varpi_{\rm Gaia}}^2)$), but different systematic error components. $\varpi_{\rm Gaia}^{\rm unc}$ (Eq. (\ref{eq:sim_unc})) has a systematic error that is simply equal to the \textit{Gaia} zero-point; while $\varpi_{\rm Gaia}^{\rm cor}$ (Eq. (\ref{eq:sim_cor})) has a systematic error centred on $\ogaia$ but also accounts for spatial correlations between the sources.

The spatially-correlated errors are assumed to be independent from the random errors, and the corresponding covariance matrix can be written as:
\begin{align}
    \mathcal{S} = E[(\varpi_{\rm i}-\ogaia)(\varpi_{\rm j}-\ogaia)] = \begin{cases}
        V_\varpi(0) & \mathrm{if\ } i=j \\ V_\varpi(\theta_{\rm ij}) & \mathrm{if\ }i\neq j
    \end{cases} \, ,
\end{align}
where $V_\varpi(\theta)$ is the spatial covariance function which solely depends on the angular distance between sources $i$ and $j$ ($\theta_{\rm ij}$). \citet{Lindegren2018} suggested
\begin{align}
    V_\varpi(\theta) \simeq (285 \, \mu\text{as}^2)\times\exp(-\theta/14^\circ)
    \label{eq:lindegren}
\end{align}
as the spatial correlation function for the systematic parallax errors. To capture the variance at the smallest scales \citep[see Fig. 14 of][]{Lindegren2018}, an additional exponential term can be added:
\begin{align}
    V_\varpi(\theta) \simeq (285 \, \mu\text{as}^2)\times\exp(-\theta/14^\circ) + (1565 \, \mu\text{as}^2)\times\exp(-\theta/0.3^\circ)\, ,
    \label{eq:brown}
\end{align}
where the number $1565$ $\mu$as$^2$ is chosen to get a total $V_\varpi(0)$ of $1850$ $\mu$as$^2$, appearing in the overview of \textit{Gaia} DR2 astrometry by Lindegren et al.\footnote{\url{https://www.cosmos.esa.int/web/gaia/dr2-known-issues} (slide 35)}. This value was obtained for quasars, with faint magnitudes ($G\geq13$); for brighter magnitudes (Cepheids), there are indications that a total $V_\varpi(0)$ of $440$ $\mu$as$^2$ would be required instead. However, owing to the uncertainty regarding the exact value that would be suitable for our sample, we prefer to be conservative by using Eq. (\ref{eq:brown}). Lastly, we also try the description of spatial covariances following \citet{Zinn2018}, namely:
\begin{align}
    V_\varpi(\theta) \simeq (135 \, \mu\text{as}^2)\times\exp(-\theta/14^\circ) \, .
    \label{eq:zinn}
\end{align}

\begin{figure*}
	\centering
	\includegraphics[width=0.8\hsize]{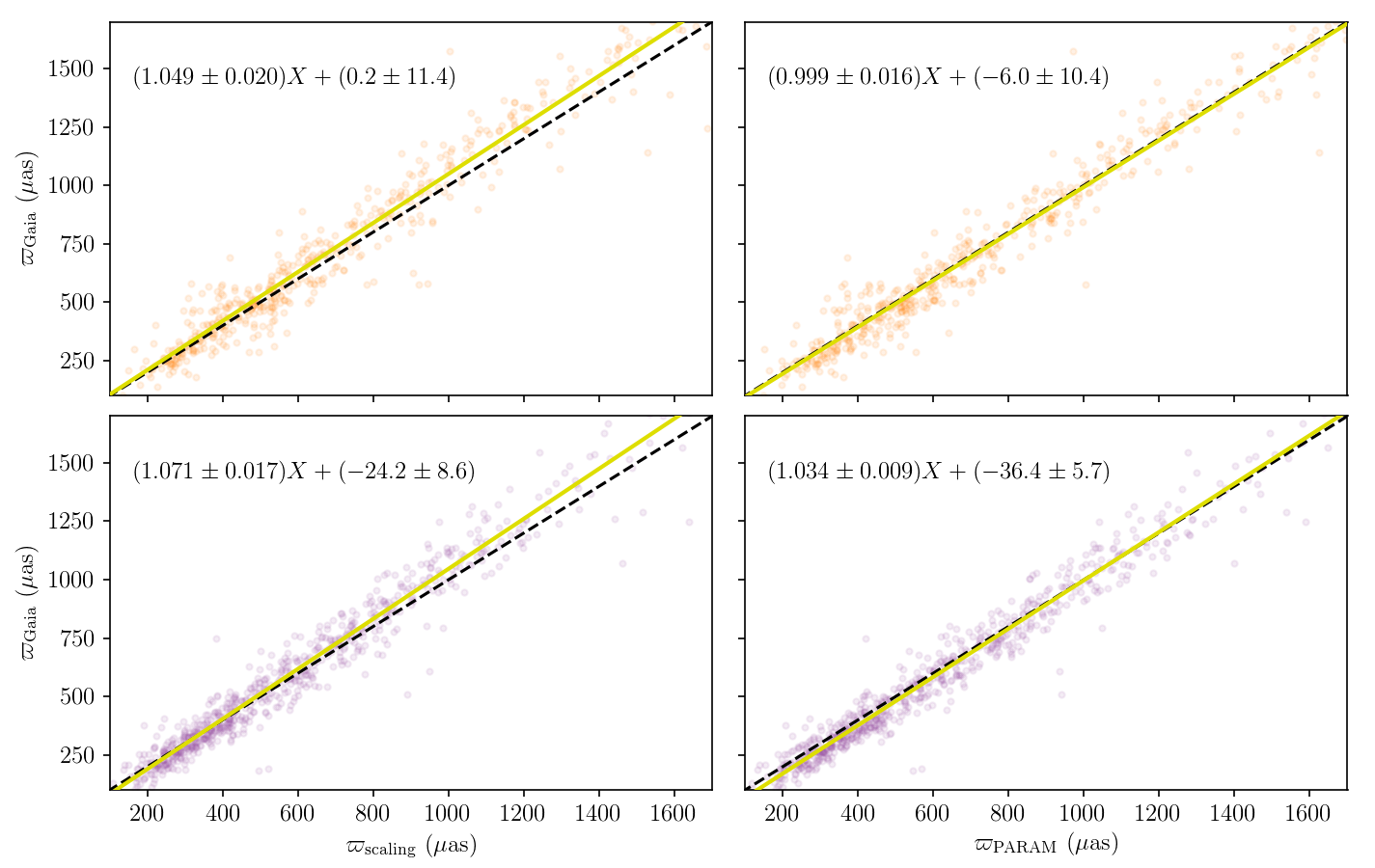}
	\caption{$\varpi_{\rm Gaia}$ as a function of $\varpi_{\rm scaling}$ (left) and $\varpi_{\rm PARAM}$ (right), for red giants in the K2 Campaign 3 (top) and 6 (bottom) fields. The yellow line displays the linear fit, averaged over $N$ realisations, for which the relation is given at the top of each subplot. The black dashed line indicates the 1:1 relation. The summary statistics are: $\left(\overline{\Delta \varpi}\right)_{\rm m} = 28.9 \pm 5.2 \ \mu$as, $\left(\overline{\Delta \varpi}\right)_{\rm w} = 24.6 \pm 4.0 \ \mu$as, and $z=1.16$ for C3 (scaling); $\left(\overline{\Delta \varpi}\right)_{\rm m} = 11.9 \pm 3.6 \ \mu$as, $\left(\overline{\Delta \varpi}\right)_{\rm w} = 9.5 \pm 2.6 \ \mu$as, and $z=1.01$ for C6 (scaling); $\left(\overline{\Delta \varpi}\right)_{\rm m} = -8.1 \pm 4.4 \ \mu$as, $\left(\overline{\Delta \varpi}\right)_{\rm w} = -6.4 \pm 3.8 \ \mu$as, and $z=1.30$ for C3 (\textsc{param}); $\left(\overline{\Delta \varpi}\right)_{\rm m} = -18.6 \pm 3.3 \ \mu$as, $\left(\overline{\Delta \varpi}\right)_{\rm w} = -16.9 \pm 2.4 \ \mu$as, and $z=1.11$ for C6 (\textsc{param}).}
	\label{fig:K2}
\end{figure*}

\begin{figure}
   \centering
   \includegraphics[width=\hsize]{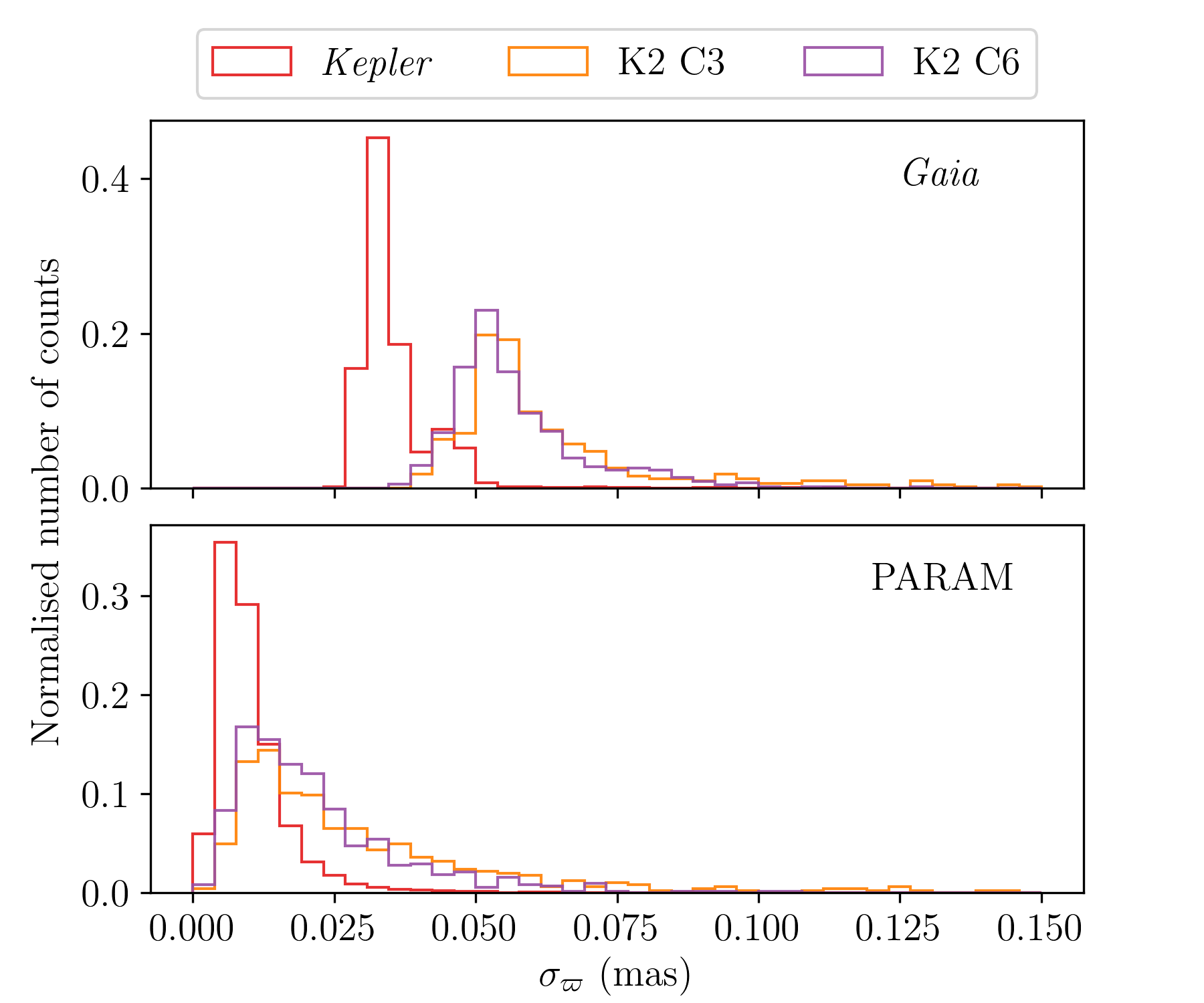}
   \caption{Distribution of the parallax uncertainties in \textit{Gaia} DR2 (top) and in \textsc{param} (bottom) for the \textit{Kepler} (red), C3 (orange), and C6 (purple) fields.}
   \label{fig:sigma_plx}
\end{figure}

As the term $\mathcal{N}(\ogaia, \mathcal{S})$ in Eq. (\ref{eq:sim_cor}) is subject to important variations between different simulations, we draw $N_{\rm sims}=1000$ realisations of $\varpi_{\rm Gaia}^{\rm cor}$ in order to obtain statistically significant results. Furthermore, because it is computationally expensive to calculate the covariance matrix for a large number of sources, we randomly select 60\,\% of the RGB and RC samples beforehand. This allows us to calculate the median parallax difference between the astrometric and seismic synthetic values, as well as its uncertainty. Whether spatial covariances are included or not, we obtain a similar offset, very close to the parallax zero-point applied ($\ogaia=-50 \ \mu$as), for both RGB and RC stars. The difference becomes apparent when one looks at the uncertainty on the median offset. Without spatial correlations, we find an uncertainty of approximately $1 \ \mu$as, which is compatible with our results. However, when spatial correlations are included, the uncertainty increases up to $\sim 14 \ \mu$as using Eqs. (\ref{eq:lindegren}) and (\ref{eq:brown}), and it is slightly lower with Eq. (\ref{eq:zinn}) ($\sim 10 \ \mu$as). A similar threshold on the uncertainty of the parallax offset, due to spatial covariances, was recently found by \citet{Hall2019}, who used hierarchical Bayesian modelling and assumptions about the red clump to compare \textit{Gaia} and asteroseismic parallaxes in the \textit{Kepler} field. Then, studying the relation between the simulated seismic and \textit{Gaia} parallaxes (in a similar way as on Fig. \ref{fig:scaling_gaia}), we find that, regardless of the spatial covariance function applied, the value and uncertainty of the slope parameter are barely affected. This is reassuring since it means that the slopes we obtained in Sects. \ref{sec:scaling_comp}, \ref{sec:rel_comp}, and \ref{sec:param_comp} are significant, and that the argument whereby \textsc{param} displays a slope closer to unity compared to the raw scaling relations is still valid. 

%__________________________________________________________________

\section{Positional dependence of the parallax zero-point}
\label{sec:position}

\begin{figure*}
   \centering
   \includegraphics[width=0.7\hsize]{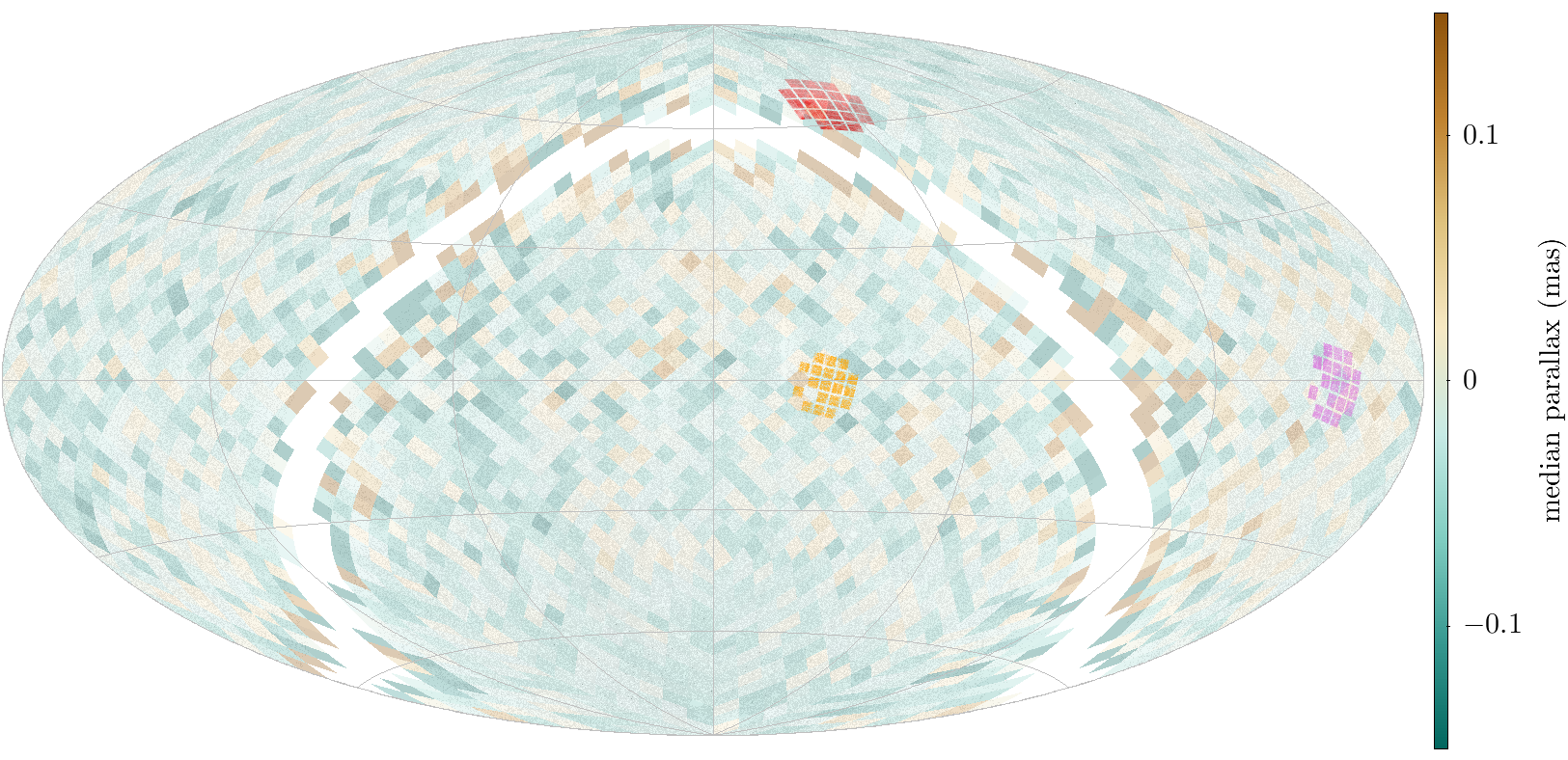}
   \caption{Sky map in ecliptic coordinates of the median parallaxes for the full quasar sample, showing large-scale variations of the parallax zero-point. The \textit{Kepler} (red), C3 (orange), and C6 (purple) fields are displayed. Median values are calculated in cells of $3.7 \times 3.7 \ \rm {deg}^2$.}
   \label{fig:map_qso}
\end{figure*}

\begin{figure}
   \centering
   \includegraphics[width=0.85\hsize]{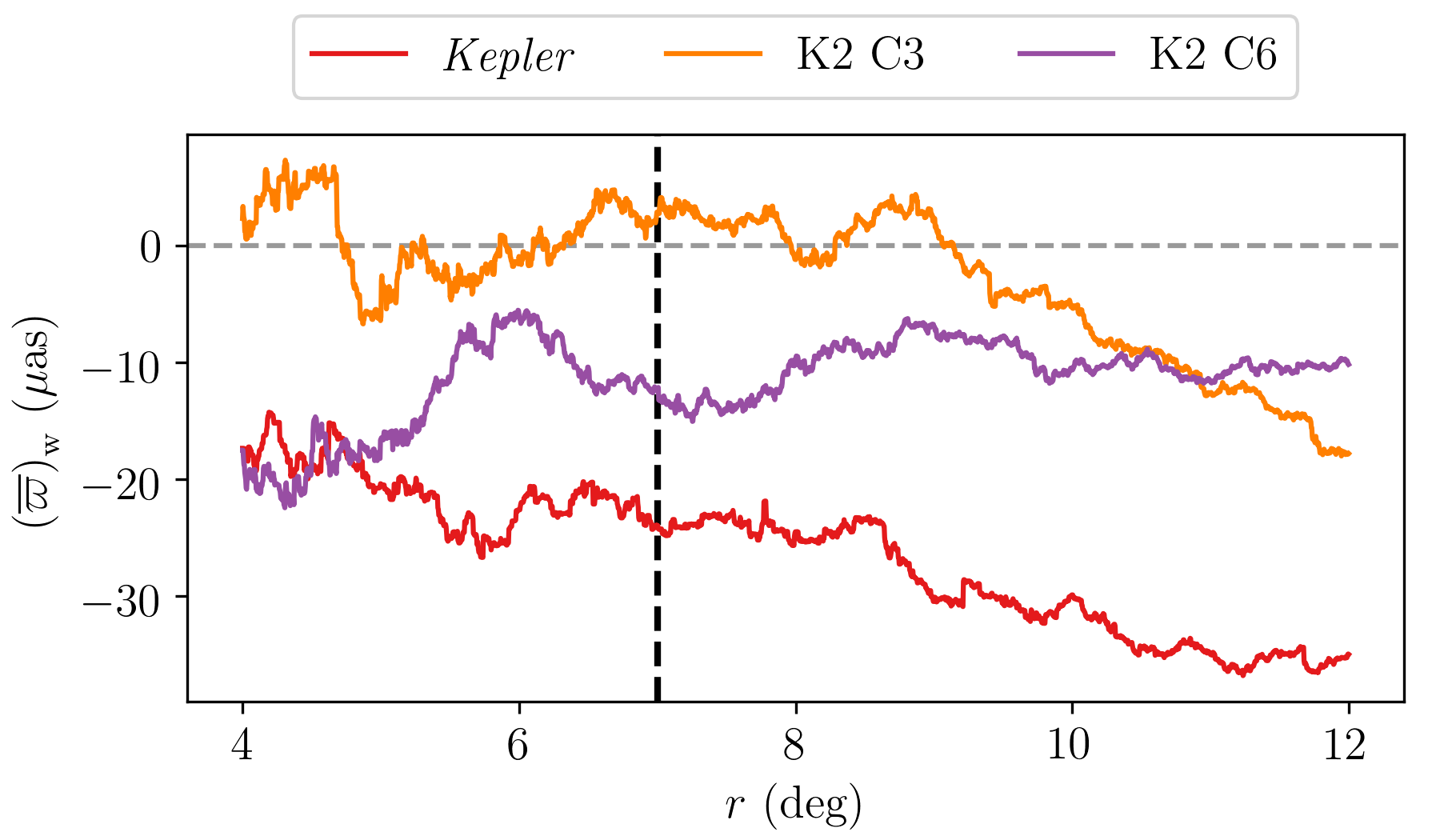}
   \caption{Weighted average parallax of the quasars selected within a given radius around the central coordinates of the \textit{Kepler} (red), C3 (orange), and C6 (purple) fields. The black dashed line indicates the average radius of the fields.}
   \label{fig:qso_plx_fields}
\end{figure}

This section aims at highlighting how the position constitutes one of the sources of the variations in the \textit{Gaia} DR2 parallax zero-point. 
After measuring the offset in the \textit{Kepler} field, we undertake a similar analysis for two of the K2 Campaign fields: C3 and C6, corresponding to the south and north Galactic caps respectively \citep{Howell2014}. A difference lies in the fact that there is no distinction between the RGB and the RC evolutionary stages here. In the following, we are concerned with the results given by raw scaling relations and \textsc{param}, using the method developed by \citet{Mosser2011} for $\deltanu$ for both as there is no $\deltanu$ available following the approach by \citet{Yu2018}. After that, we also analyse the information given by quasars regarding the three fields considered in our investigation.

\subsection{K2 fields: C3 and C6}

The comparison of parallaxes using the raw scalings with parallaxes from \textit{Gaia} DR2 is displayed on Fig. \ref{fig:K2}, and the measured offsets are of the order of $25 \pm 4$ and $9 \pm 3 \ \mu$as for C3 and C6, respectively. Both have a $\varpi_{\rm scaling}$---$\varpi_{\rm Gaia}$ relation with a slope substantially different from unity, i.e. $1.049 \pm 0.020$ for C3 and $1.071 \pm 0.017$ for C6. 
Besides, $T_{\rm eff}$ shifts of $\pm 100 \ \rm K$ affect the parallax difference by $\pm$ 10-15 $\mu$as, as is the case for \textit{Kepler}.

For \textsc{param}, the outcome of the comparison is also illustrated on Fig. \ref{fig:K2}. C3 displays a parallax difference close to zero ($\Delta \varpi \simeq -6 \pm 4 \ \mu$as), while C6 shows a value of about $-17 \pm 2 \ \mu$as. In the case of C3, the trend with parallax is entirely suppressed: the slope is equal to $0.999 \pm 0.016$. It is also reduced for C6, but a fairly steep slope of $1.034 \pm 0.009$ remains. 
In absolute terms, these offsets are much lower compared to the \textit{Kepler} field even though we are dealing with red-giant stars, either in the RGB or the RC phase, in both cases. Thus, in the position-magnitude-colour dependence of the parallax zero-point, the position prevails in the current 
analysis. 
As to whether these differences are caused by the inhomogeneity in the effective temperatures and metallicities in use between the \textit{Kepler} and K2 samples, \citet{Casagrande2019} checked the reliability of their photometric metallicities against APOGEE DR14 and found an offset of $-0.01$ dex with an RMS of 0.25 dex, i.e. SkyMapper's [Fe/H] are lower. $T_{\rm eff}$ from SkyMapper agree with APOGEE DR14 within few tens of K and a typical RMS of 100 K. These small deviations should not affect our findings. Also, we compare the extinctions from \textsc{param}, adopted in this work, to those from SkyMapper, which are used to determine $T_{\rm eff}$ and [Fe/H], and find that they are consistent with each other, with differences in $A_K$ only at the level of $\pm 0.01$. 
Finally, we investigate the potential origin of the differing slopes between C3 and C6, by considering the different parallax distributions (with respect to each other, and also to the \textit{Kepler} field), the decreased quality of the K2 seismic data compared to \textit{Kepler} \citep{Howell2014}, and the use of $\deltanu$ from \citet{Mosser2011} \citep[instead of][as for \textit{Kepler}]{Yu2018}. Nevertheless, the interplay between these different elements as well as the limited number of stars in the K2 samples prevent us from drawing any firm conclusion with regards to the slopes. As for the offsets, the effect of such differences seems to be at the level of $\sim \pm 7 \ \mu$as at most. In addition, the fairly good agreement between SkyMapper and APOGEE does not exclude the possibility that the photometric $T_{\rm eff}$ and [Fe/H] may be influenced differently in different fields (e.g. C3 and C6) by external factors such as, e.g., the extinction.

A further point we would like to raise concerns the apparent larger scatter in the K2 fields. In that respect, we identify the order of magnitude of the \textit{Gaia} and seismic (\textsc{param}) parallax uncertainties (Fig. \ref{fig:sigma_plx}). The asteroseismic uncertainties are slightly larger in the case of K2. This can mainly be explained by the fact that the original mission, \textit{Kepler}, continuously monitored stars for four years, whereas each campaign of K2 is limited to a duration of approximately 80 days \citep{Howell2014}. Despite this, photometric systematics in the K2 data leading to, e.g., spurious frequencies are now well understood, and the pipelines used to produce lightcurves have been developed over time to remove or mitigate these effects. As a result, we do not expect the global seismic parameters to show any significant bias due to the K2 artefacts \citep[see also][and the \textit{Kepler} and K2 Science Center website\footnote{\url{https://keplerscience.arc.nasa.gov/pipeline.html}}]{Hekker2012}. Nonetheless, the astrometric uncertainties are also substantially larger for K2, almost doubled compared to \textit{Kepler}. A possible reason for this would be that the regions around the ecliptic plane (such as C3 and C6) are observed less frequently, as a result of the \textit{Gaia} scanning law, and also under less favourable scanning geometry, with the scan angles not distributed evenly \citep{GaiaCollaboration2016a}. To test this hypothesis, we use the quantity \texttt{visibility\_periods\_used}: the number of visibility periods, i.e. a group of observations separated from other groups by a gap of at least 4 days, used in the astrometric solution. This way, we can assess if a source is astrometrically well-observed. This variable exhibits significantly higher values for \textit{Kepler}, ranging from 12 to 17. The number of visibility periods for K2 are lower or equal to ten, indicating that the parallaxes could be more vulnerable to errors. The predicted uncertainty contrast between the \textit{Kepler} and K2 fields is about a factor\footnote{\url{https://www.cosmos.esa.int/web/gaia/science-performance}} of 1.6, which is indeed consistent with the location of the histogram peaks in the top panel of Fig. \ref{fig:sigma_plx}. 

\subsection{Quasars and Colour-Magnitude Diagram}

\begin{figure*}
   \centering
   \includegraphics[width=0.75\hsize]{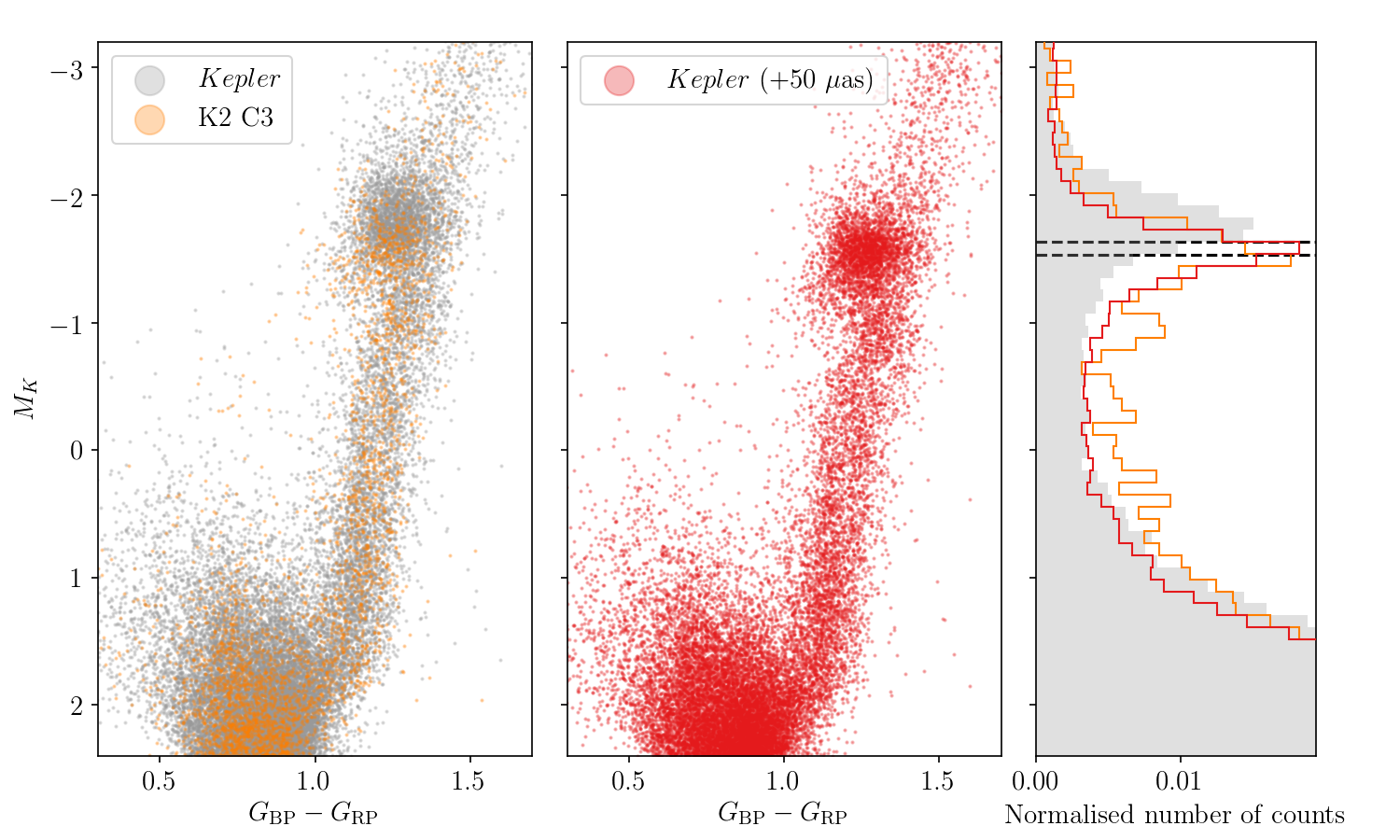}
   \caption{Colour-Magnitude Diagrams (CMDs; left and middle) and absolute magnitude $M_K$ normalised histograms (right), where $M_K$ is estimated by means of the \textit{Gaia} parallax at face value for the \textit{Kepler} (grey) and C3 (orange) fields. Another CMD, including a shift in parallax, is shown for \textit{Kepler} (red). We removed stars having a parallax with a relative error above 10\,\% for \textit{Kepler}, and 15\,\% for C3. The black dashed lines indicate the expected range of values for the magnitude of the clump in the $K$ band.}
   \label{fig:cmds_plxshift}
\end{figure*}

\citet{Lindegren2018} investigated the parallax offset using quasars, and obtained a global zero-point of about $-30 \ \mu$as. It is no surprise that this value differs from the ones we obtain in the current work. Indeed, this parallax offset depends on magnitude and colour, in addition to position: quasars generally are blue-coloured with faint magnitudes. Red giants are substantially different objects, hence the importance to solve the parallax zero-point independently. We investigate the information provided by quasars to estimate the parallax zero-point in the different fields considered here (see Fig. \ref{fig:map_qso}). To this end, we select quasars within a given radius around the central coordinates of each field and compute the weighted average parallax (following Eq. (\ref{eq:wmean})). The variation of this quantity with radius is shown on Fig. \ref{fig:qso_plx_fields}. This allows us to assess its sensitivity on the size of the region considered. The mean offsets 
associated to the size of the fields ($r \sim 7^\circ$) are $-24\pm8$, $3\pm9$, and $-12\pm7 \ \mu$as for \textit{Kepler}, C3, and C6, respectively. It should be kept in mind, however, that spatial covariances in the parallax errors (Sect. \ref{sec:covariances}) prevent one from drawing strong conclusions regarding the offsets, especially at the smallest scales, and the main purpose of Fig. \ref{fig:qso_plx_fields} is to illustrate the trends. Despite exhibiting different values, the pattern whereby the \textit{Gaia} parallax offset is lowest for C3 and highest for \textit{Kepler}, in keeping with our results, is potentially reproduced.

For illustrative purposes, we show on Fig. \ref{fig:cmds_plxshift} two Colour-Magnitude Diagrams (CMDs) for the entire \textit{Kepler} field: one where the absolute magnitude is calculated without applying a shift in the \textit{Gaia} parallax, another one where we use the zero-point measured with \textsc{param} ($\sim -50 \ \mu$as for both RGB and RC stars) to ``correct'' the parallaxes. In addition, owing to the near-zero offset in C3, the latter is also displayed for comparison. We only keep stars with a relative parallax error below 10\,\% for \textit{Kepler}, and 15\,\% for C3. Beyond the magnitude values being affected, the shape of the RGB structures, e.g. the red-giant branch bump and the red clump, become clearer when a shift is applied. This is a sensible change, because a constant shift in parallax is not equivalent to a constant shift in luminosity. The parallax shift has a different relative effect on each star's distance, hence luminosity, which may explain how features in the CMD can become sharper. In particular, the red clump becomes more sharply-peaked in the absolute magnitude distribution and its mean value is about $M_{K}^{\rm RC} \sim -1.57$, versus $M_{K}^{\rm RC} \sim -1.78$ when \textit{Gaia} parallaxes are taken at face value. Independent determinations of $M_{K}^{\rm RC}$ range between $-1.63$ and $-1.53$ \citep[see Table 1 of][]{Girardi2016,Chen2017,Hawkins2017}. Furthermore, population effects at a level of several hundredths of a magnitude are expected but are not enough to explain the difference in $M_{K}^{\rm RC}$, especially considering that the use of the $K$ band partly mitigates them \citep[see, e.g.,][and references therein]{Girardi2016}.

%__________________________________________________________________

\section{Joint calibration of the seismic scaling relations and of the zero-point in the \textit{Gaia} parallaxes}
\label{sec:calibration}

In Sect. \ref{sec:scaling_comp}, we used the scaling relations (Eqs. (\ref{eq:mass_scaling}) and (\ref{eq:rad_scaling})) at face value in the context of a comparison with \textit{Gaia} DR2. These relations are not precisely calibrated yet, and testing their validity has been a very active topic in the field of asteroseismology \citep[e.g.][]{Huber2012,Miglio2012,Gaulme2016,Sahlholdt2018}. In this vein, \textit{Gaia} DR2 ensures the continuity of the research effort carried out to test the scaling relations' accuracy. After the work conducted in Sect. \ref{sec:kepler} and \ref{sec:position}, it is clear that the scalings' calibration by means of \textit{Gaia} requires the parallax zero-point to be characterised at the same time. Hence, the current investigation reflects two main developments: constraining the calibration of the seismic scaling relations and quantifying the parallax offset in \textit{Gaia} DR2.

\begin{figure}
   \centering
   \includegraphics[width=\hsize]{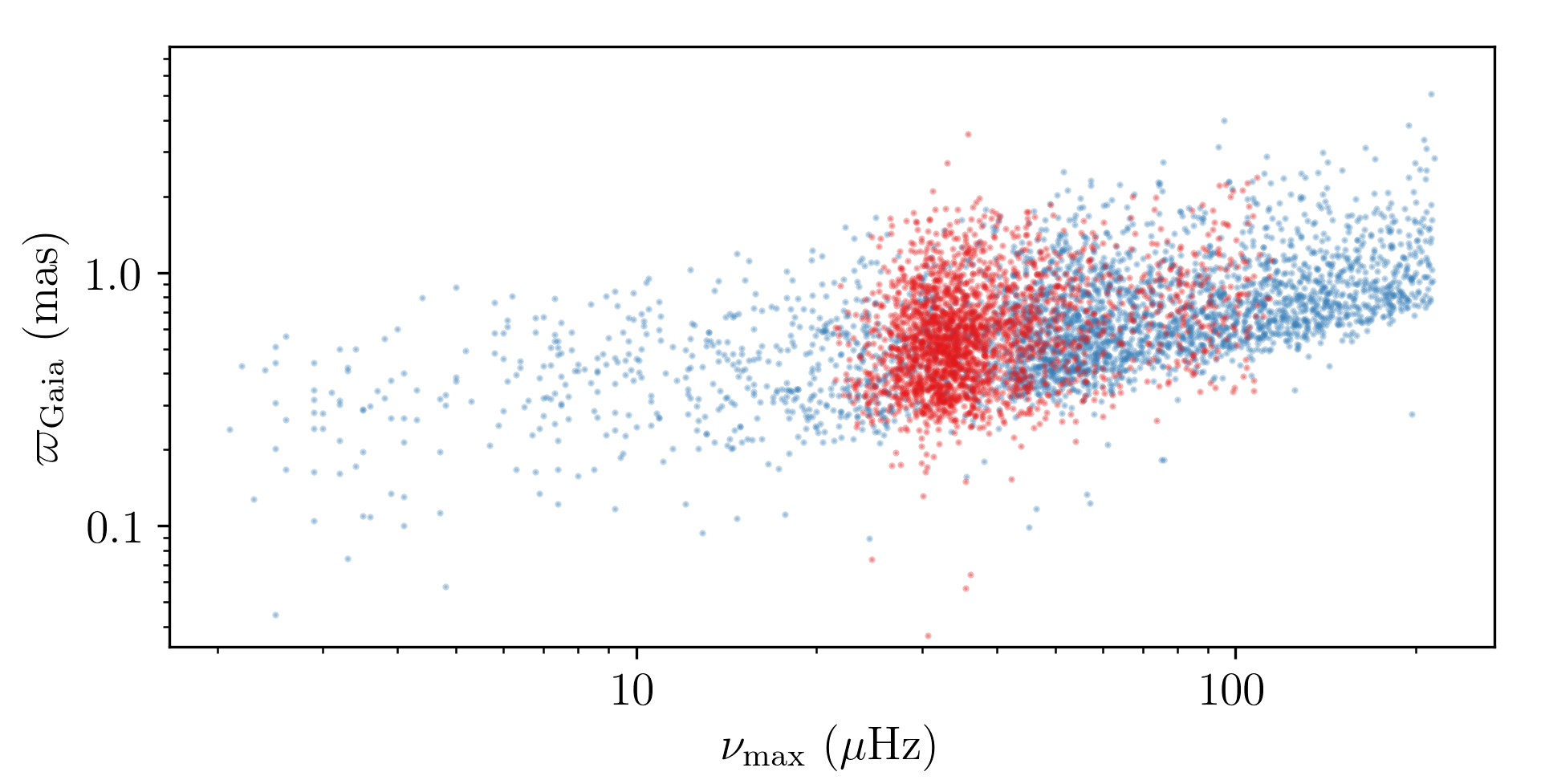}
   \caption{$\varpi_{Gaia}$ as a function of $\nu_{\rm max}$ for RGB (blue) and RC (red) stars in the \textit{Kepler} sample.}
   \label{fig:numax_Gaiaplx}
\end{figure}

\begin{figure*}
   \centering
   \includegraphics[width=0.49\hsize]{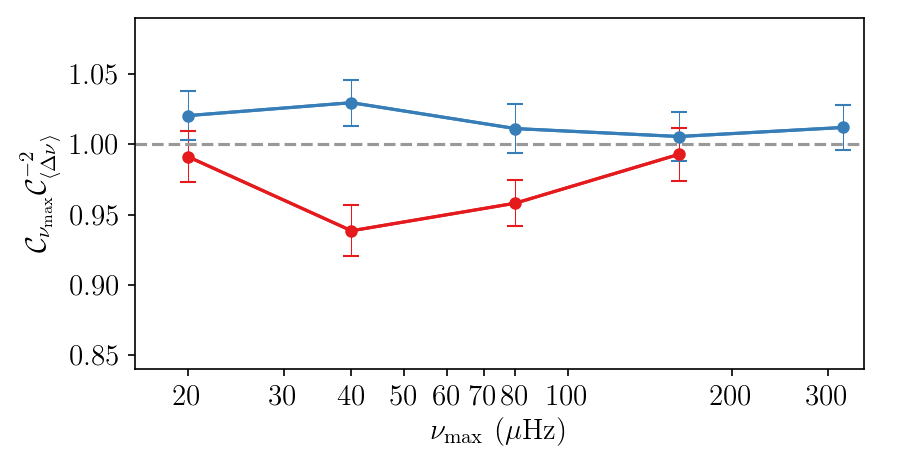}
   \includegraphics[width=0.49\hsize]{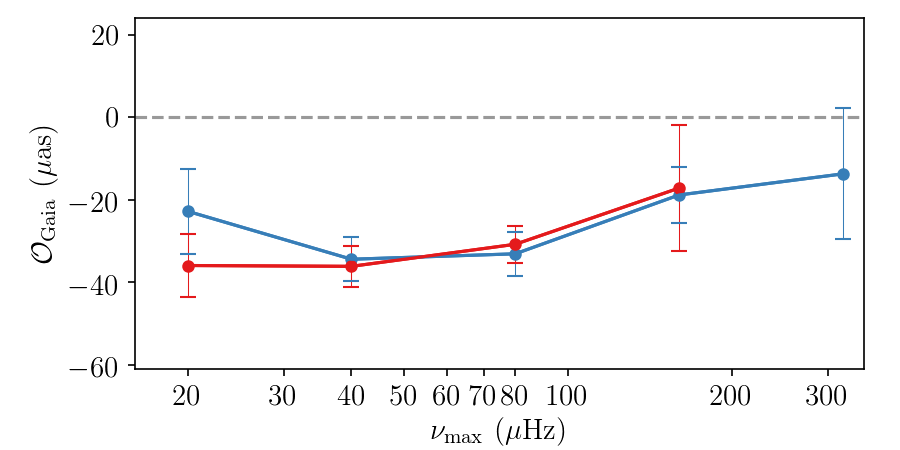}
   \caption{Coefficients (left) and offsets (right) determined via our two-step calibration methodology for RGB (blue) and RC (red) stars, as a function of $\numax$.}
   \label{fig:coef_off}
\end{figure*}

We take the seismic calibration issue into account by introducing the scaling factors $\caldnu$ and $\calnum$ in the expressions of $\deltanu$ and $\nu_{\rm max}$ (Eqs. (\ref{eq:Dnu}) and (\ref{eq:numax})):
\begin{align}
	\left( \frac{\deltanu}{\deltanu_{\rm \odot}} \right) &= \caldnu \left(\frac{M}{M_{\rm \odot}} \right)^{1/2} \left(\frac{R}{R_{\rm \odot}} \right)^{-3/2} \, , \label{eq:Dnu_calib} \\
	\left( \frac{\nu_{\rm max}}{\nu_{\rm max,\odot}} \right) &= \calnum \left(\frac{M}{M_{\rm \odot}} \right) \left(\frac{R}{R_{\rm \odot}} \right)^{-2} \left(\frac{T_{\rm eff}}{T_{\rm eff,\odot}} \right)^{-1/2} \, . \label{eq:numax_calib}
\end{align}
In terms of mass and radius (Eqs. (\ref{eq:mass_scaling}) and (\ref{eq:rad_scaling})), this translates into:
\begin{align}
    \left( \frac{M}{M_{\rm \odot}} \right) &= \calnum^{-3} \caldnu^4 \left(\frac{\nu_{\rm max}}{\nu_{\rm max, \odot}} \right)^3 \left(\frac{\deltanu}{\deltanu_{\rm \odot}} \right)^{-4} \left(\frac{T_{\rm eff}}{T_{\rm eff,\odot}} \right)^{3/2} \, , \label{eq:M_calib} \\
	\left( \frac{R}{R_{\rm \odot}} \right) &= \calnum^{-1} \caldnu^2 \left(\frac{\nu_{\rm max}}{\nu_{\rm max, \odot}} \right) \left(\frac{\deltanu}{\deltanu_{\rm \odot}} \right)^{-2} \left(\frac{T_{\rm eff}}{T_{\rm eff,\odot}} \right)^{1/2} \, . \label{eq:R_calib}
\end{align}
Finally, the seismic parallax (Eq. (\ref{eq:plx_scaling})) is modified as follows:
\begin{align}
	\varpi_{\rm scaling}' = c_{\lambda} \, \calnum \caldnu^{-2} \left(\frac{\nu_{\rm max}}{\nu_{\rm max, \odot}} \right)^{-1} \left(\frac{\deltanu}{\deltanu_{\rm \odot}} \right)^2 \left(\frac{T_{\rm eff}}{T_{\rm eff, \odot}} \right)^{-5/2} \, . \label{eq:plx_calib}
\end{align}

In this context, the comparison between the \textit{Gaia} (Eq. (\ref{eq:plx_Gaia})) and asteroseismic (Eq. (\ref{eq:plx_calib})) expressions for the parallax gives the following equality
\begin{align}
	\varpi_{\rm Gaia} - \ogaia &= \calnum \caldnu^{-2} \varpi_{\rm scaling} \, , \label{eq:comp_scaling}
\end{align}
where $\ogaia$ represents the parallax zero-point in \textit{Gaia} DR2 to be determined. Fitting Eq. (\ref{eq:comp_scaling}) with a \textsc{ransac} algorithm allows us to determine the coefficient $\calnum \caldnu^{-2}$, accounting for the scaling relations' calibration, and also provides an offset $\ogaia$, which we can interpret as a bias in the \textit{Gaia} parallaxes. The main assumption that we make here is that the asteroseismic calibration, in the form of a multiplication factor, and the astrometric calibration, in the form of an addition factor, can be considered independently and do not affect each other. For this reason, it is crucial to make efficient use of both asteroseismic and astrometric data. On the one hand, because corrections to the scaling relations are expected to depend on $\numax$ \citep[see, e.g., Fig. 3 of][]{Rodrigues2017}, we divide our \textit{Kepler} RGB and RC samples in frequency ranges of $\numax$ values: [8, 32], [16, 64], [32, 128], [64, 256], [128, 512] $\mu$Hz. On the other hand, nearby stars have more reliable parallaxes (less affected, in relative terms, by the \textit{Gaia} offset) and may, as such, be used to calibrate the scalings. In practice, we implement a two-step methodology to, firstly, calibrate the seismic scaling relations and, secondly, use the calibration coefficients obtained from the first step to determine the \textit{Gaia} zero-point. To do so, we start by selecting stars with large parallaxes in each bin of $\numax$. As illustrated by Fig. \ref{fig:numax_Gaiaplx}, the high-parallax threshold has to be chosen differently depending on the $\numax$ bin considered, in order to keep enough stars. Here, the limit is chosen in such a way that at least 500 stars remain in the different $\numax$ ranges. We then interpolate in $\numax$ to estimate the scaling factors and individually correct each seismic parallax. The latter is then compared again to the \textit{Gaia} parallax, this time on the full range of parallaxes, to measure the parallax offset.

During the calibration process, we apply linear fits expressed in the following form: $\varpi_{\rm Gaia} = \gamma \varpi_{\rm scaling} + \delta$, where $\gamma = \calnum \caldnu^{-2}$ is the calibration parameter and $\delta = \ogaia $ is the offset parameter. The parameters' uncertainties are estimated by repeating \textsc{ransac} $N=1000$ times, where we add a normally distributed noise knowing the observed uncertainties on $\varpi_{\rm Gaia}$ and $\varpi_{\rm scaling}$. The coefficients, obtained in step 1, and offsets, obtained in step 2, are shown as a function of $\numax$ for RGB and RC stars on Fig. \ref{fig:coef_off}. 
The offsets $\ogaia$ point in the right direction --- \textit{Gaia} parallaxes are smaller --- and are in the same order of magnitude for the two evolutionary stages, which validates the calibration of the scaling relations. These offsets do not depend on $\numax$, and their mean values are $-24 \pm 9 \ \mu$as for RGB stars and $-31 \pm 7 \ \mu$as for RC stars. 
In regard to the scaling factors, we can make a qualitative comparison with the $\caldnu$ estimated from models if we assume $\calnum$ to be equal to unity \citep[uncertainties related to modelling the driving and damping of oscillations prevented theoretical tests of the $\numax$ scaling relation; see, e.g.,][]{Belkacem2011}. According to \citet{Rodrigues2017} (see their Fig. 3), $\caldnu$ takes values slighly lower / higher than one for RGB / RC stars in the ranges of mass and metallicity concerning our sample. Additionally, from RGB models, $\caldnu$ is expected to decrease before increasing again as we go towards increasing $\numax$ values, with a minimum at $\numax \sim 15 \ \mu$Hz depending on mass and metallicity. A similar trend is expected for RC stars but the other way around: $\caldnu$ increases before decreasing, and has a maximum at $\numax \sim 30 \ \mu$Hz which depends again on $M$ and [Fe/H]. Also, larger variations of $\caldnu$ are expected for RGB stars, which seems in conflict with our findings. Nevertheless, we remind that we derive scaling factors that are averaged in bins of $\numax$, and it appears that the current results are still in general agreement with expectations from models. 
At this point, the third Data Release of \textit{Gaia}, coming along with smaller uncertainties, will provide the means to pursue this work, and to derive precise and accurate corrections to the scaling relations. 

%__________________________________________________________________

\section{Conclusions}
\label{sec:conclusions}

We combined \textit{Gaia} and \textit{Kepler} data to investigate the \textit{Gaia} DR2 parallax zero-point, showing how the measured offsets depend on the asteroseismic method employed, having a direct illustration of the positional dependence of the zero-point thanks to the K2 fields, and, finally, introducing a way to address the seismic and astrometric calibrations at the same time.

First of all, 
the application of three distinct asteroseismic methods, in the course of the comparison with the astrometric parallaxes delivered by the second Data Release of \textit{Gaia}, reveals that there is no absolute standard within asteroseismology. The determination of a zero-point in the \textit{Gaia} parallaxes extensively depends on the seismic approach used, and cannot be dissociated from it. As a matter of fact, the conclusions we draw are not the same whether we use the seismic scaling relations at face value or a grid-based method such as \textsc{param}: the former would suggest a near-zero offset for RGB stars, significantly different from that of RC stars; in contrast, the latter implies a similar deviation with respect to \textit{Gaia} DR2 parallaxes for RGB and RC stars. That said, the offsets measured when using \textsc{param}, ranging from $\sim -45$ to $-55 \ \mu$as --- considering the substantial uncertainties induced by spatially-correlated errors --- can be related to previous investigations, especially the one conducted by \citet{Zinn2018} who calibrated seismic radii against eclipsing binary data in clusters and used model-predicted corrections to the $\deltanu$ scaling relation. 
They obtained very similar offsets of about $-50 \ \mu$as for RGB and RC stars observed by \textit{Kepler}. Our external validation via the measurements from eclipsing binaries in the open clusters NGC 6791 and NGC 6819 also confirms the existence of a parallax offset in that range. The proximity of the \textsc{param} results with these independent tests 
reaffirms previous findings about the necessity to go beyond the $\deltanu$ scaling for the estimation of stellar properties. 
Furthermore, the use of different sets of $\deltanu$ values has a non-negligible impact on the inferred offsets, of the order of $\sim 10 \ \mu$as. In particular, attention should be paid to the consistency in the definition of $\deltanu$ between the observations and the models. Other systematic effects can arise from, e.g., shifts in the effective temperature and metallicity scales, and changes in the physical inputs of the models, with variations up to $\pm 7 \ \mu$as according to our tests but very likely larger than that due to uncertainties related to stellar models.

We also bring to light the positional dependence of the \textit{Gaia} DR2 parallax zero-point, as demonstrated by \citet{Lindegren2018}, by analysing two of the K2 Campaign fields, C3 and C6, in addition to the \textit{Kepler} field. These fields, corresponding to the south and north Galactic caps, display parallax offsets which are substantially different from \textit{Kepler}'s. A small fraction of these differences may be due, e.g, to the parallax distribution, the quality of the seismic data, and the use of different seismic constraints. But, as of now, it remains difficult to reach a firm conclusion, and the future possibility of extending this analysis to other K2 Campaign fields may help shedding light on this matter. Also, despite the measured values being slightly different, the offset suggested by quasars reproduces the trend towards the increasing discrepancy with \textit{Gaia} for C3, C6, and \textit{Kepler} (in ascending order). The difference in the calculated zero-point is to be expected because quasars have their own peculiarities (e.g. faint magnitude, blue colour) that are not representative of red-giant stars. Furthermore, such a colour dependence also emerges from the study of $\delta$ Scuti stars in the \textit{Kepler} field by \citet{Murphy2019}, who found that applying an offset (as high as 30 $\mu$as) resulted in unrealistically low luminosities. Looking forward, having a uniform set of spectroscopic constraints would be very valuable.

Lastly, we initiate a two-step model-independent method to simultaneously calibrate the asteroseismic scaling relations and measure the \textit{Gaia} DR2 parallax zero-point, based on the assumption that these two corrections are fully decoupled. This leads us to promising findings whereby the computed calibration coefficients are qualitatively comparable to those that are derived from models, and the estimated offsets are in the same order of magnitude for RGB and RC stars and suggest that \textit{Gaia} parallaxes are too small --- as expected. However, given the non-negligible uncertainties and the close correlation between the calibration and offset parameters, it is still too soon to draw strong conclusions. In this regard, the third Data Release of \textit{Gaia}, with improved parallax uncertainties and reduced systematics, will offer exciting prospects to continue along the path of calibrating the scaling relations.

\begin{acknowledgements}
This work has made use of data from the European Space Agency (ESA) mission
{\it Gaia} (\url{https://www.cosmos.esa.int/gaia}), processed by the {\it Gaia} Data Processing and Analysis Consortium (DPAC, \url{https://www.cosmos.esa.int/web/gaia/dpac/consortium}). Funding for the DPAC has been provided by national institutions, in particular the institutions participating in the {\it Gaia} Multilateral Agreement.
SK, AM, BM, GRD, BMR, DB, and LG are grateful to the International Space Science Institute (ISSI) for support provided to the \mbox{asteroSTEP} ISSI International Team. AM, WJC, GRD, BMR, YPE, and TSHN acknowledge the support of the UK Science and Technology Facilities Council (STFC). AM acknowledges support from the ERC Consolidator Grant funding scheme ({\em project ASTEROCHRONOMETRY}, G.A. n. 772293). LC is the recipient of the ARC Future Fellowship FT160100402. TSR acknowledges financial support from Premiale 2015 MITiC (PI B. Garilli). This work was supported by FCT/MCTES through national funds and by FEDER - Fundo Europeu de Desenvolvimento Regional through COMPETE2020 - Programa Operacional Competitividade e Internacionalização by these grants: UID/FIS/04434/2019; PTDC/FIS-AST/30389/2017 \& POCI-01-0145-FEDER-030389. DB is supported in the form of work contract funded by national funds through Fundação para a Ciência e Tecnologia (FCT). We also wish to thank the referee whose comments helped clarify the paper and interpret further the results.
\end{acknowledgements}

\bibliographystyle{aa} % style aa.bst
\bibliography{references} % your references Yourfile.bib

\begin{thebibliography}{73}
\expandafter\ifx\csname natexlab\endcsname\relax\def\natexlab#1{#1}\fi

\bibitem[{{Abolfathi} {et~al.}(2018){Abolfathi}, {Aguado}, {Aguilar}, {Allende
  Prieto}, {Almeida}, {Tasnim Ananna}, {Anders}, {Anderson}, {Andrews},
  {Anguiano}, \& et~al.}]{Abolfathi2018}
{Abolfathi}, B., {Aguado}, D.~S., {Aguilar}, G., {et~al.} 2018, \apjs, 235, 42

\bibitem[{{Arenou} {et~al.}(2018){Arenou}, {Luri}, {Babusiaux}, {Fabricius},
  {Helmi}, {Muraveva}, {Robin}, {Spoto}, {Vallenari}, {Antoja},
  {Cantat-Gaudin}, {Jordi}, {Leclerc}, {Reyl{\'e}}, {Romero-G{\'o}mez}, {Shih},
  {Soria}, {Barache}, {Bossini}, {Bragaglia}, {Breddels}, {Fabrizio},
  {Lambert}, {Marrese}, {Massari}, {Moitinho}, {Robichon}, {Ruiz-Dern},
  {Sordo}, {Veljanoski}, {Eyer}, {Jasniewicz}, {Pancino}, {Soubiran}, {Spagna},
  {Tanga}, {Turon}, \& {Zurbach}}]{Arenou2018}
{Arenou}, F., {Luri}, X., {Babusiaux}, C., {et~al.} 2018, \aap, 616, A17

\bibitem[{{Basu} {et~al.}(2010){Basu}, {Chaplin}, \& {Elsworth}}]{Basu2010}
{Basu}, S., {Chaplin}, W.~J., \& {Elsworth}, Y. 2010, \apj, 710, 1596

\bibitem[{{Belkacem}(2012)}]{Belkacem2012}
{Belkacem}, K. 2012, in SF2A-2012: Proceedings of the Annual meeting of the
  French Society of Astronomy and Astrophysics, ed. S.~{Boissier}, P.~{de
  Laverny}, N.~{Nardetto}, R.~{Samadi}, D.~{Valls-Gabaud}, \& H.~{Wozniak},
  173--188

\bibitem[{{Belkacem} {et~al.}(2011){Belkacem}, {Goupil}, {Dupret}, {Samadi},
  {Baudin}, {Noels}, \& {Mosser}}]{Belkacem2011}
{Belkacem}, K., {Goupil}, M.~J., {Dupret}, M.~A., {et~al.} 2011, \aap, 530,
  A142

\bibitem[{{Belkacem} {et~al.}(2013){Belkacem}, {Samadi}, {Mosser}, {Goupil}, \&
  {Ludwig}}]{Belkacem2013}
{Belkacem}, K., {Samadi}, R., {Mosser}, B., {Goupil}, M.-J., \& {Ludwig}, H.-G.
  2013, in Astronomical Society of the Pacific Conference Series, Vol. 479,
  Progress in Physics of the Sun and Stars: A New Era in Helio- and
  Asteroseismology, ed. H.~{Shibahashi} \& A.~E. {Lynas-Gray}, 61

\bibitem[{{Brogaard} {et~al.}(2011){Brogaard}, {Bruntt}, {Grundahl}, {Clausen},
  {Frandsen}, {Vandenberg}, \& {Bedin}}]{Brogaard2011}
{Brogaard}, K., {Bruntt}, H., {Grundahl}, F., {et~al.} 2011, \aap, 525, A2

\bibitem[{{Brogaard} {et~al.}(2018){Brogaard}, {Hansen}, {Miglio}, {Slumstrup},
  {Frandsen}, {Jessen-Hansen}, {Lund}, {Bossini}, {Thygesen}, {Davies},
  {Chaplin}, {Arentoft}, {Bruntt}, {Grundahl}, \& {Handberg}}]{Brogaard2018}
{Brogaard}, K., {Hansen}, C.~J., {Miglio}, A., {et~al.} 2018, \mnras, 476, 3729

\bibitem[{{Brogaard} {et~al.}(2016){Brogaard}, {Jessen-Hansen}, {Handberg},
  {Arentoft}, {Frandsen}, {Grundahl}, {Bruntt}, {Sandquist}, {Miglio}, {Beck},
  {Thygesen}, {Kj{\ae}rgaard}, \& {Haugaard}}]{Brogaard2016}
{Brogaard}, K., {Jessen-Hansen}, J., {Handberg}, R., {et~al.} 2016,
  Astronomische Nachrichten, 337, 793

\bibitem[{{Brogaard} {et~al.}(2012){Brogaard}, {VandenBerg}, {Bruntt},
  {Grundahl}, {Frandsen}, {Bedin}, {Milone}, {Dotter}, {Feiden}, {Stetson},
  {Sandquist}, {Miglio}, {Stello}, \& {Jessen-Hansen}}]{Brogaard2012}
{Brogaard}, K., {VandenBerg}, D.~A., {Bruntt}, H., {et~al.} 2012, \aap, 543,
  A106

\bibitem[{{Brown} {et~al.}(1991){Brown}, {Gilliland}, {Noyes}, \&
  {Ramsey}}]{Brown1991}
{Brown}, T.~M., {Gilliland}, R.~L., {Noyes}, R.~W., \& {Ramsey}, L.~W. 1991,
  \apj, 368, 599

\bibitem[{{Cantat-Gaudin} {et~al.}(2018){Cantat-Gaudin}, {Jordi}, {Vallenari},
  {Bragaglia}, {Balaguer-N{\'u}{\~n}ez}, {Soubiran}, {Bossini}, {Moitinho},
  {Castro-Ginard}, {Krone-Martins}, {Casamiquela}, {Sordo}, \&
  {Carrera}}]{Cantat-Gaudin2018}
{Cantat-Gaudin}, T., {Jordi}, C., {Vallenari}, A., {et~al.} 2018, \aap, 618,
  A93

\bibitem[{{Casagrande} \& {VandenBerg}(2014)}]{Casagrande2014a}
{Casagrande}, L. \& {VandenBerg}, D.~A. 2014, \mnras, 444, 392

\bibitem[{{Casagrande} \& {VandenBerg}(2018{\natexlab{a}})}]{Casagrande2018a}
{Casagrande}, L. \& {VandenBerg}, D.~A. 2018{\natexlab{a}}, \mnras, 479, L102

\bibitem[{{Casagrande} \& {VandenBerg}(2018{\natexlab{b}})}]{Casagrande2018}
{Casagrande}, L. \& {VandenBerg}, D.~A. 2018{\natexlab{b}}, \mnras, 475, 5023

\bibitem[{{Casagrande} {et~al.}(2019){Casagrande}, {Wolf}, {Mackey},
  {Nordlander}, {Yong}, \& {Bessell}}]{Casagrande2019}
{Casagrande}, L., {Wolf}, C., {Mackey}, A.~D., {et~al.} 2019, \mnras, 482, 2770

\bibitem[{{Chen} {et~al.}(2017){Chen}, {Casagrande}, {Zhao}, {Bovy}, {Silva
  Aguirre}, {Zhao}, \& {Jia}}]{Chen2017}
{Chen}, Y.~Q., {Casagrande}, L., {Zhao}, G., {et~al.} 2017, \apj, 840, 77

\bibitem[{{Christensen-Dalsgaard}(2002)}]{Christensen-Dalsgaard2002}
{Christensen-Dalsgaard}, J. 2002, Reviews of Modern Physics, 74, 1073

\bibitem[{{Davies} {et~al.}(2017){Davies}, {Lund}, {Miglio}, {Elsworth},
  {Kuszlewicz}, {North}, {Rendle}, {Chaplin}, {Rodrigues}, {Campante},
  {Girardi}, {Hale}, {Hall}, {Jones}, {Kawaler}, {Roxburgh}, \&
  {Schofield}}]{Davies2017}
{Davies}, G.~R., {Lund}, M.~N., {Miglio}, A., {et~al.} 2017, \aap, 598, L4

\bibitem[{{Davies} {et~al.}(2016){Davies}, {Silva Aguirre}, {Bedding},
  {Handberg}, {Lund}, {Chaplin}, {Huber}, {White}, {Benomar}, {Hekker}, {Basu},
  {Campante}, {Christensen-Dalsgaard}, {Elsworth}, {Karoff}, {Kjeldsen},
  {Lundkvist}, {Metcalfe}, \& {Stello}}]{Davies2016a}
{Davies}, G.~R., {Silva Aguirre}, V., {Bedding}, T.~R., {et~al.} 2016, \mnras,
  456, 2183

\bibitem[{{De Ridder} {et~al.}(2016){De Ridder}, {Molenberghs}, {Eyer}, \&
  {Aerts}}]{DeRidder2016}
{De Ridder}, J., {Molenberghs}, G., {Eyer}, L., \& {Aerts}, C. 2016, \aap, 595,
  L3

\bibitem[{{Elsworth} {et~al.}(2017){Elsworth}, {Hekker}, {Basu}, \&
  {Davies}}]{Elsworth2017}
{Elsworth}, Y., {Hekker}, S., {Basu}, S., \& {Davies}, G.~R. 2017, \mnras, 466,
  3344

\bibitem[{{Farmer} {et~al.}(2013){Farmer}, {Kolb}, \& {Norton}}]{Farmer2013}
{Farmer}, R., {Kolb}, U., \& {Norton}, A.~J. 2013, \mnras, 433, 1133

\bibitem[{Fischler \& Bolles(1981)}]{Fischler1981}
Fischler, M.~A. \& Bolles, R.~C. 1981, Commun. ACM, 24, 381

\bibitem[{{Gai} {et~al.}(2011){Gai}, {Basu}, {Chaplin}, \&
  {Elsworth}}]{Gai2011}
{Gai}, N., {Basu}, S., {Chaplin}, W.~J., \& {Elsworth}, Y. 2011, \apj, 730, 63

\bibitem[{{Gaia Collaboration} {et~al.}(2018){Gaia Collaboration}, {Brown},
  {Vallenari}, {Prusti}, {de Bruijne}, {Babusiaux}, {Bailer-Jones}, {Biermann},
  {Evans}, {Eyer}, \& et~al.}]{GaiaCollaboration2018}
{Gaia Collaboration}, {Brown}, A.~G.~A., {Vallenari}, A., {et~al.} 2018, \aap,
  616, A1

\bibitem[{{Gaia Collaboration} {et~al.}(2016{\natexlab{a}}){Gaia
  Collaboration}, {Brown}, {Vallenari}, {Prusti}, {de Bruijne}, {Mignard},
  {Drimmel}, {Babusiaux}, {Bailer-Jones}, {Bastian}, \&
  et~al.}]{GaiaCollaboration2016}
{Gaia Collaboration}, {Brown}, A.~G.~A., {Vallenari}, A., {et~al.}
  2016{\natexlab{a}}, \aap, 595, A2

\bibitem[{{Gaia Collaboration} {et~al.}(2016{\natexlab{b}}){Gaia
  Collaboration}, {Prusti}, {de Bruijne}, {Brown}, {Vallenari}, {Babusiaux},
  {Bailer-Jones}, {Bastian}, {Biermann}, {Evans}, {Eyer}, {Jansen}, {Jordi},
  {Klioner}, {Lammers}, {Lindegren}, {Luri}, {Mignard}, {Milligan}, {Panem},
  {Poinsignon}, {Pourbaix}, {Randich}, {Sarri}, {Sartoretti}, {Siddiqui},
  {Soubiran}, {Valette}, {van Leeuwen}, {Walton}, {Aerts}, {Arenou}, {Cropper},
  {Drimmel}, {H{\o}g}, {Katz}, {Lattanzi}, {O'Mullane}, {Grebel}, {Holland},
  {Huc}, {Passot}, {Bramante}, {Cacciari}, {Casta{\~n}eda}, {Chaoul}, {Cheek},
  {De Angeli}, {Fabricius}, {Guerra}, {Hern{\'a}ndez}, {Jean-Antoine-Piccolo},
  {Masana}, {Messineo}, {Mowlavi}, {Nienartowicz}, {Ord{\'o}{\~n}ez- Blanco},
  {Panuzzo}, {Portell}, {Richards}, {Riello}, {Seabroke}, {Tanga},
  {Th{\'e}venin}, {Torra}, {Els}, {Gracia- Abril}, {Comoretto},
  {Garcia-Reinaldos}, {Lock}, {Mercier}, {Altmann}, {Andrae}, {Astraatmadja},
  {Bellas-Velidis}, {Benson}, {Berthier}, {Blomme}, {Busso}, {Carry},
  {Cellino}, {Clementini}, {Cowell}, {Creevey}, {Cuypers}, {Davidson}, {De
  Ridder}, {de Torres}, {Delchambre}, {Dell'Oro}, {Ducourant}, {Fr{\'e}mat},
  {Garc{\'\i}a-Torres}, {Gosset}, {Halbwachs}, {Hambly}, {Harrison}, {Hauser},
  {Hestroffer}, {Hodgkin}, {Huckle}, {Hutton}, {Jasniewicz}, {Jordan},
  {Kontizas}, {Korn}, {Lanzafame}, {Manteiga}, {Moitinho}, {Muinonen},
  {Osinde}, {Pancino}, {Pauwels}, {Petit}, {Recio-Blanco}, {Robin}, {Sarro},
  {Siopis}, {Smith}, {Smith}, {Sozzetti}, {Thuillot}, {van Reeven}, {Viala},
  {Abbas}, {Abreu Aramburu}, {Accart}, {Aguado}, {Allan}, {Allasia},
  {Altavilla}, {{\'A}lvarez}, {Alves}, {Anderson}, {Andrei}, {Anglada Varela},
  {Antiche}, {Antoja}, {Ant{\'o}n}, {Arcay}, {Atzei}, {Ayache}, {Bach},
  {Baker}, {Balaguer-N{\'u}{\~n}ez}, {Barache}, {Barata}, {Barbier}, {Barblan},
  {Baroni}, {Barrado y Navascu{\'e}s}, {Barros}, {Barstow}, {Becciani},
  {Bellazzini}, {Bellei}, {Bello Garc{\'\i}a}, {Belokurov}, {Bendjoya},
  {Berihuete}, {Bianchi}, {Bienaym{\'e}}, {Billebaud}, {Blagorodnova},
  {Blanco-Cuaresma}, {Boch}, {Bombrun}, {Borrachero}, {Bouquillon}, {Bourda},
  {Bouy}, {Bragaglia}, {Breddels}, {Brouillet}, {Br{\"u}semeister},
  {Bucciarelli}, {Budnik}, {Burgess}, {Burgon}, {Burlacu}, {Busonero}, {Buzzi},
  {Caffau}, {Cambras}, {Campbell}, {Cancelliere}, {Cantat-Gaudin}, {Carlucci},
  {Carrasco}, {Castellani}, {Charlot}, {Charnas}, {Charvet}, {Chassat},
  {Chiavassa}, {Clotet}, {Cocozza}, {Collins}, {Collins}, {Costigan}, {Crifo},
  {Cross}, {Crosta}, {Crowley}, {Dafonte}, {Damerdji}, {Dapergolas}, {David},
  {David}, {De Cat}, {de Felice}, {de Laverny}, {De Luise}, {De March}, {de
  Martino}, {de Souza}, {Debosscher}, {del Pozo}, {Delbo}, {Delgado},
  {Delgado}, {di Marco}, {Di Matteo}, {Diakite}, {Distefano}, {Dolding}, {Dos
  Anjos}, {Drazinos}, {Dur{\'a}n}, {Dzigan}, {Ecale}, {Edvardsson}, {Enke},
  {Erdmann}, {Escolar}, {Espina}, {Evans}, {Eynard Bontemps}, {Fabre},
  {Fabrizio}, {Faigler}, {Falc{\~a}o}, {Farr{\`a}s Casas}, {Faye}, {Federici},
  {Fedorets}, {Fern{\'a}ndez-Hern{\'a}ndez}, {Fernique}, {Fienga}, {Figueras},
  {Filippi}, {Findeisen}, {Fonti}, {Fouesneau}, {Fraile}, {Fraser}, {Fuchs},
  {Furnell}, {Gai}, {Galleti}, {Galluccio}, {Garabato}, {Garc{\'\i}a-Sedano},
  {Gar{\'e}}, {Garofalo}, {Garralda}, {Gavras}, {Gerssen}, {Geyer}, {Gilmore},
  {Girona}, {Giuffrida}, {Gomes}, {Gonz{\'a}lez-Marcos},
  {Gonz{\'a}lez-N{\'u}{\~n}ez}, {Gonz{\'a}lez-Vidal}, {Granvik}, {Guerrier},
  {Guillout}, {Guiraud}, {G{\'u}rpide}, {Guti{\'e}rrez-S{\'a}nchez}, {Guy},
  {Haigron}, {Hatzidimitriou}, {Haywood}, {Heiter}, {Helmi}, {Hobbs},
  {Hofmann}, {Holl}, {Holland}, {Hunt}, {Hypki}, {Icardi}, {Irwin}, {Jevardat
  de Fombelle}, {Jofr{\'e}}, {Jonker}, {Jorissen}, {Julbe}, {Karampelas},
  {Kochoska}, {Kohley}, {Kolenberg}, {Kontizas}, {Koposov}, {Kordopatis},
  {Koubsky}, {Kowalczyk}, {Krone-Martins}, {Kudryashova}, {Kull}, {Bachchan},
  {Lacoste-Seris}, {Lanza}, {Lavigne}, {Le Poncin-Lafitte}, {Lebreton},
  {Lebzelter}, {Leccia}, {Leclerc}, {Lecoeur-Taibi}, {Lemaitre}, {Lenhardt},
  {Leroux}, {Liao}, {Licata}, {Lindstr{\o}m}, {Lister}, {Livanou}, {Lobel},
  {L{\"o}ffler}, {L{\'o}pez}, {Lopez-Lozano}, {Lorenz}, {Loureiro},
  {MacDonald}, {Magalh{\~a}es Fernandes}, {Managau}, {Mann}, {Mantelet},
  {Marchal}, {Marchant}, {Marconi}, {Marie}, {Marinoni}, {Marrese},
  {Marschalk{\'o}}, {Marshall}, {Mart{\'\i}n-Fleitas}, {Martino}, {Mary},
  {Matijevi{\v{c}}}, {Mazeh}, {McMillan}, {Messina}, {Mestre}, {Michalik},
  {Millar}, {Miranda}, {Molina}, {Molinaro}, {Molinaro}, {Moln{\'a}r},
  {Moniez}, {Montegriffo}, {Monteiro}, {Mor}, {Mora}, {Morbidelli}, {Morel},
  {Morgenthaler}, {Morley}, {Morris}, {Mulone}, {Muraveva}, {Musella},
  {Narbonne}, {Nelemans}, {Nicastro}, {Noval}, {Ord{\'e}novic},
  {Ordieres-Mer{\'e}}, {Osborne}, {Pagani}, {Pagano}, {Pailler}, {Palacin},
  {Palaversa}, {Parsons}, {Paulsen}, {Pecoraro}, {Pedrosa}, {Pentik{\"a}inen},
  {Pereira}, {Pichon}, {Piersimoni}, {Pineau}, {Plachy}, {Plum}, {Poujoulet},
  {Pr{\v{s}}a}, {Pulone}, {Ragaini}, {Rago}, {Rambaux}, {Ramos-Lerate},
  {Ranalli}, {Rauw}, {Read}, {Regibo}, {Renk}, {Reyl{\'e}}, {Ribeiro},
  {Rimoldini}, {Ripepi}, {Riva}, {Rixon}, {Roelens}, {Romero-G{\'o}mez},
  {Rowell}, {Royer}, {Rudolph}, {Ruiz-Dern}, {Sadowski}, {Sagrist{\`a}
  Sell{\'e}s}, {Sahlmann}, {Salgado}, {Salguero}, {Sarasso}, {Savietto},
  {Schnorhk}, {Schultheis}, {Sciacca}, {Segol}, {Segovia}, {Segransan},
  {Serpell}, {Shih}, {Smareglia}, {Smart}, {Smith}, {Solano}, {Solitro},
  {Sordo}, {Soria Nieto}, {Souchay}, {Spagna}, {Spoto}, {Stampa}, {Steele},
  {Steidelm{\"u}ller}, {Stephenson}, {Stoev}, {Suess}, {S{\"u}veges}, {Surdej},
  {Szabados}, {Szegedi-Elek}, {Tapiador}, {Taris}, {Tauran}, {Taylor},
  {Teixeira}, {Terrett}, {Tingley}, {Trager}, {Turon}, {Ulla}, {Utrilla},
  {Valentini}, {van Elteren}, {Van Hemelryck}, {van Leeuwen}, {Varadi},
  {Vecchiato}, {Veljanoski}, {Via}, {Vicente}, {Vogt}, {Voss}, {Votruba},
  {Voutsinas}, {Walmsley}, {Weiler}, {Weingrill}, {Werner}, {Wevers},
  {Whitehead}, {Wyrzykowski}, {Yoldas}, {{\v{Z}}erjal}, {Zucker}, {Zurbach},
  {Zwitter}, {Alecu}, {Allen}, {Allende Prieto}, {Amorim},
  {Anglada-Escud{\'e}}, {Arsenijevic}, {Azaz}, {Balm}, {Beck}, {Bernstein},
  {Bigot}, {Bijaoui}, {Blasco}, {Bonfigli}, {Bono}, {Boudreault}, {Bressan},
  {Brown}, {Brunet}, {Bunclark}, {Buonanno}, {Butkevich}, {Carret}, {Carrion},
  {Chemin}, {Ch{\'e}reau}, {Corcione}, {Darmigny}, {de Boer}, {de Teodoro}, {de
  Zeeuw}, {Delle Luche}, {Domingues}, {Dubath}, {Fodor}, {Fr{\'e}zouls},
  {Fries}, {Fustes}, {Fyfe}, {Gallardo}, {Gallegos}, {Gardiol}, {Gebran},
  {Gomboc}, {G{\'o}mez}, {Grux}, {Gueguen}, {Heyrovsky}, {Hoar}, {Iannicola},
  {Isasi Parache}, {Janotto}, {Joliet}, {Jonckheere}, {Keil}, {Kim},
  {Klagyivik}, {Klar}, {Knude}, {Kochukhov}, {Kolka}, {Kos}, {Kutka}, {Lainey},
  {LeBouquin}, {Liu}, {Loreggia}, {Makarov}, {Marseille}, {Martayan},
  {Martinez-Rubi}, {Massart}, {Meynadier}, {Mignot}, {Munari}, {Nguyen},
  {Nordlander}, {Ocvirk}, {O'Flaherty}, {Olias Sanz}, {Ortiz}, {Osorio},
  {Oszkiewicz}, {Ouzounis}, {Palmer}, {Park}, {Pasquato}, {Peltzer}, {Peralta},
  {P{\'e}turaud}, {Pieniluoma}, {Pigozzi}, {Poels}, {Prat}, {Prod'homme},
  {Raison}, {Rebordao}, {Risquez}, {Rocca-Volmerange}, {Rosen}, {Ruiz-Fuertes},
  {Russo}, {Sembay}, {Serraller Vizcaino}, {Short}, {Siebert}, {Silva},
  {Sinachopoulos}, {Slezak}, {Soffel}, {Sosnowska}, {Strai{\v{z}}ys}, {ter
  Linden}, {Terrell}, {Theil}, {Tiede}, {Troisi}, {Tsalmantza}, {Tur},
  {Vaccari}, {Vachier}, {Valles}, {Van Hamme}, {Veltz}, {Virtanen}, {Wallut},
  {Wichmann}, {Wilkinson}, {Ziaeepour}, \& {Zschocke}}]{GaiaCollaboration2016a}
{Gaia Collaboration}, {Prusti}, T., {de Bruijne}, J.~H.~J., {et~al.}
  2016{\natexlab{b}}, \aap, 595, A1

\bibitem[{{Gaulme} {et~al.}(2016){Gaulme}, {McKeever}, {Jackiewicz}, {Rawls},
  {Corsaro}, {Mosser}, {Southworth}, {Mahadevan}, {Bender}, \&
  {Deshpande}}]{Gaulme2016}
{Gaulme}, P., {McKeever}, J., {Jackiewicz}, J., {et~al.} 2016, \apj, 832, 121

\bibitem[{Girardi(2016)}]{Girardi2016}
Girardi, L. 2016, Annual Review of Astronomy and Astrophysics, 54, 95

\bibitem[{{Girardi} {et~al.}(2002){Girardi}, {Bertelli}, {Bressan}, {Chiosi},
  {Groenewegen}, {Marigo}, {Salasnich}, \& {Weiss}}]{Girardi2002}
{Girardi}, L., {Bertelli}, G., {Bressan}, A., {et~al.} 2002, \aap, 391, 195

\bibitem[{{Green} {et~al.}(2015){Green}, {Schlafly}, {Finkbeiner}, {Rix},
  {Martin}, {Burgett}, {Draper}, {Flewelling}, {Hodapp}, {Kaiser}, {Kudritzki},
  {Magnier}, {Metcalfe}, {Price}, {Tonry}, \& {Wainscoat}}]{Green2015}
{Green}, G.~M., {Schlafly}, E.~F., {Finkbeiner}, D.~P., {et~al.} 2015, \apj,
  810, 25

\bibitem[{{Guggenberger} {et~al.}(2016){Guggenberger}, {Hekker}, {Basu}, \&
  {Bellinger}}]{Guggenberger2016}
{Guggenberger}, E., {Hekker}, S., {Basu}, S., \& {Bellinger}, E. 2016, \mnras,
  460, 4277

\bibitem[{{Hall} {et~al.}(2019){Hall}, {Davies}, {Elsworth}, {Miglio},
  {Bedding}, {Brown}, {Khan}, {Hawkins}, {Garc{\'\i}a}, {Chaplin}, \&
  {North}}]{Hall2019}
{Hall}, O.~J., {Davies}, G.~R., {Elsworth}, Y.~P., {et~al.} 2019, \mnras, 486,
  3569

\bibitem[{{Handberg} {et~al.}(2017){Handberg}, {Brogaard}, {Miglio}, {Bossini},
  {Elsworth}, {Slumstrup}, {Davies}, \& {Chaplin}}]{Handberg2017}
{Handberg}, R., {Brogaard}, K., {Miglio}, A., {et~al.} 2017, \mnras, 472, 979

\bibitem[{{Hawkins} {et~al.}(2017){Hawkins}, {Leistedt}, {Bovy}, \&
  {Hogg}}]{Hawkins2017}
{Hawkins}, K., {Leistedt}, B., {Bovy}, J., \& {Hogg}, D.~W. 2017, \mnras, 471,
  722

\bibitem[{{Hekker} {et~al.}(2012){Hekker}, {Elsworth}, {Mosser}, {Kallinger},
  {Chaplin}, {De Ridder}, {Garc{\'\i}a}, {Stello}, {Clarke}, {Hall}, \&
  {Ibrahim}}]{Hekker2012}
{Hekker}, S., {Elsworth}, Y., {Mosser}, B., {et~al.} 2012, \aap, 544, A90

\bibitem[{{Howell} {et~al.}(2014){Howell}, {Sobeck}, {Haas}, {Still},
  {Barclay}, {Mullally}, {Troeltzsch}, {Aigrain}, {Bryson}, {Caldwell},
  {Chaplin}, {Cochran}, {Huber}, {Marcy}, {Miglio}, {Najita}, {Smith},
  {Twicken}, \& {Fortney}}]{Howell2014}
{Howell}, S.~B., {Sobeck}, C., {Haas}, M., {et~al.} 2014, \pasp, 126, 398

\bibitem[{{Huber} {et~al.}(2012){Huber}, {Ireland}, {Bedding}, {Brand{\~a}o},
  {Piau}, {Maestro}, {White}, {Bruntt}, {Casagrande}, {Molenda-{\.Z}akowicz},
  {Silva Aguirre}, {Sousa}, {Barclay}, {Burke}, {Chaplin},
  {Christensen-Dalsgaard}, {Cunha}, {De Ridder}, {Farrington}, {Frasca},
  {Garc{\'{\i}}a}, {Gilliland}, {Goldfinger}, {Hekker}, {Kawaler}, {Kjeldsen},
  {McAlister}, {Metcalfe}, {Miglio}, {Monteiro}, {Pinsonneault}, {Schaefer},
  {Stello}, {Stumpe}, {Sturmann}, {Sturmann}, {ten Brummelaar}, {Thompson},
  {Turner}, \& {Uytterhoeven}}]{Huber2012}
{Huber}, D., {Ireland}, M.~J., {Bedding}, T.~R., {et~al.} 2012, \apj, 760, 32

\bibitem[{{Huber} {et~al.}(2017){Huber}, {Zinn}, {Bojsen-Hansen},
  {Pinsonneault}, {Sahlholdt}, {Serenelli}, {Silva Aguirre}, {Stassun},
  {Stello}, {Tayar}, {Bastien}, {Bedding}, {Buchhave}, {Chaplin}, {Davies},
  {Garc{\'{\i}}a}, {Latham}, {Mathur}, {Mosser}, \& {Sharma}}]{Huber2017}
{Huber}, D., {Zinn}, J., {Bojsen-Hansen}, M., {et~al.} 2017, \apj, 844, 102

\bibitem[{{Kjeldsen} \& {Bedding}(1995)}]{Kjeldsen1995}
{Kjeldsen}, H. \& {Bedding}, T.~R. 1995, \aap, 293, 87

\bibitem[{{Lagarde} {et~al.}(2015){Lagarde}, {Miglio}, {Eggenberger}, {Morel},
  {Montalb{\'a}n}, {Mosser}, {Rodrigues}, {Girardi}, {Rainer}, {Poretti},
  {Barban}, {Hekker}, {Kallinger}, {Valentini}, {Carrier}, {Hareter},
  {Mantegazza}, {Elsworth}, {Michel}, \& {Baglin}}]{Lagarde2015}
{Lagarde}, N., {Miglio}, A., {Eggenberger}, P., {et~al.} 2015, \aap, 580, A141

\bibitem[{{Lindegren} {et~al.}(2018){Lindegren}, {Hern{\'a}ndez}, {Bombrun},
  {Klioner}, {Bastian}, {Ramos-Lerate}, {de Torres}, {Steidelm{\"u}ller},
  {Stephenson}, {Hobbs}, {Lammers}, {Biermann}, {Geyer}, {Hilger}, {Michalik},
  {Stampa}, {McMillan}, {Casta{\~n}eda}, {Clotet}, {Comoretto}, {Davidson},
  {Fabricius}, {Gracia}, {Hambly}, {Hutton}, {Mora}, {Portell}, {van Leeuwen},
  {Abbas}, {Abreu}, {Altmann}, {Andrei}, {Anglada}, {Balaguer-N{\'u}{\~n}ez},
  {Barache}, {Becciani}, {Bertone}, {Bianchi}, {Bouquillon}, {Bourda},
  {Br{\"u}semeister}, {Bucciarelli}, {Busonero}, {Buzzi}, {Cancelliere},
  {Carlucci}, {Charlot}, {Cheek}, {Crosta}, {Crowley}, {de Bruijne}, {de
  Felice}, {Drimmel}, {Esquej}, {Fienga}, {Fraile}, {Gai}, {Garralda},
  {Gonz{\'a}lez-Vidal}, {Guerra}, {Hauser}, {Hofmann}, {Holl}, {Jordan},
  {Lattanzi}, {Lenhardt}, {Liao}, {Licata}, {Lister}, {L{\"o}ffler},
  {Marchant}, {Martin-Fleitas}, {Messineo}, {Mignard}, {Morbidelli}, {Poggio},
  {Riva}, {Rowell}, {Salguero}, {Sarasso}, {Sciacca}, {Siddiqui}, {Smart},
  {Spagna}, {Steele}, {Taris}, {Torra}, {van Elteren}, {van Reeven}, \&
  {Vecchiato}}]{Lindegren2018}
{Lindegren}, L., {Hern{\'a}ndez}, J., {Bombrun}, A., {et~al.} 2018, \aap, 616,
  A2

\bibitem[{{Lindegren} {et~al.}(2016){Lindegren}, {Lammers}, {Bastian},
  {Hern{\'a}ndez}, {Klioner}, {Hobbs}, {Bombrun}, {Michalik}, {Ramos-Lerate},
  {Butkevich}, {Comoretto}, {Joliet}, {Holl}, {Hutton}, {Parsons},
  {Steidelm{\"u}ller}, {Abbas}, {Altmann}, {Andrei}, {Anton}, {Bach},
  {Barache}, {Becciani}, {Berthier}, {Bianchi}, {Biermann}, {Bouquillon},
  {Bourda}, {Br{\"u}semeister}, {Bucciarelli}, {Busonero}, {Carlucci},
  {Casta{\~n}eda}, {Charlot}, {Clotet}, {Crosta}, {Davidson}, {de Felice},
  {Drimmel}, {Fabricius}, {Fienga}, {Figueras}, {Fraile}, {Gai}, {Garralda},
  {Geyer}, {Gonz{\'a}lez-Vidal}, {Guerra}, {Hambly}, {Hauser}, {Jordan},
  {Lattanzi}, {Lenhardt}, {Liao}, {L{\"o}ffler}, {McMillan}, {Mignard}, {Mora},
  {Morbidelli}, {Portell}, {Riva}, {Sarasso}, {Serraller}, {Siddiqui}, {Smart},
  {Spagna}, {Stampa}, {Steele}, {Taris}, {Torra}, {van Reeven}, {Vecchiato},
  {Zschocke}, {de Bruijne}, {Gracia}, {Raison}, {Lister}, {Marchant},
  {Messineo}, {Soffel}, {Osorio}, {de Torres}, \& {O'Mullane}}]{Lindegren2016}
{Lindegren}, L., {Lammers}, U., {Bastian}, U., {et~al.} 2016, \aap, 595, A4

\bibitem[{{Luri} {et~al.}(2018){Luri}, {Brown}, {Sarro}, {Arenou},
  {Bailer-Jones}, {Castro-Ginard}, {de Bruijne}, {Prusti}, {Babusiaux}, \&
  {Delgado}}]{Luri2018}
{Luri}, X., {Brown}, A.~G.~A., {Sarro}, L.~M., {et~al.} 2018, \aap, 616, A9

\bibitem[{{Majewski} {et~al.}(2011){Majewski}, {Zasowski}, \&
  {Nidever}}]{Majewski2011}
{Majewski}, S.~R., {Zasowski}, G., \& {Nidever}, D.~L. 2011, \apj, 739, 25

\bibitem[{{Miglio}(2012)}]{Miglio2012}
{Miglio}, A. 2012, Astrophysics and Space Science Proceedings, 26, 11

\bibitem[{{Miglio} {et~al.}(2012){Miglio}, {Brogaard}, {Stello}, {Chaplin},
  {D'Antona}, {Montalb{\'a}n}, {Basu}, {Bressan}, {Grundahl}, {Pinsonneault},
  {Serenelli}, {Elsworth}, {Hekker}, {Kallinger}, {Mosser}, {Ventura},
  {Bonanno}, {Noels}, {Silva Aguirre}, {Szabo}, {Li}, {McCauliff}, {Middour},
  \& {Kjeldsen}}]{Miglio2012a}
{Miglio}, A., {Brogaard}, K., {Stello}, D., {et~al.} 2012, \mnras, 419, 2077

\bibitem[{{Miglio} {et~al.}(2016){Miglio}, {Chaplin}, {Brogaard}, {Lund},
  {Mosser}, {Davies}, {Handberg}, {Milone}, {Marino}, {Bossini}, {Elsworth},
  {Grundahl}, {Arentoft}, {Bedin}, {Campante}, {Jessen-Hansen}, {Jones},
  {Kuszlewicz}, {Malavolta}, {Nascimbeni}, \& {Sandquist}}]{Miglio2016}
{Miglio}, A., {Chaplin}, W.~J., {Brogaard}, K., {et~al.} 2016, \mnras, 461, 760

\bibitem[{{Miglio} {et~al.}(2010){Miglio}, {Montalb{\'a}n}, {Carrier}, {De
  Ridder}, {Mosser}, {Eggenberger}, {Scuflaire}, {Ventura}, {D'Antona},
  {Noels}, \& {Baglin}}]{Miglio2010}
{Miglio}, A., {Montalb{\'a}n}, J., {Carrier}, F., {et~al.} 2010, \aap, 520, L6

\bibitem[{{Mosser} {et~al.}(2011){Mosser}, {Belkacem}, {Goupil}, {Michel},
  {Elsworth}, {Barban}, {Kallinger}, {Hekker}, {De Ridder}, {Samadi}, {Baudin},
  {Pinheiro}, {Auvergne}, {Baglin}, \& {Catala}}]{Mosser2011}
{Mosser}, B., {Belkacem}, K., {Goupil}, M.~J., {et~al.} 2011, \aap, 525, L9

\bibitem[{{Murphy} {et~al.}(2019){Murphy}, {Hey}, {Van Reeth}, \&
  {Bedding}}]{Murphy2019}
{Murphy}, S.~J., {Hey}, D., {Van Reeth}, T., \& {Bedding}, T.~R. 2019, \mnras,
  485, 2380

\bibitem[{{Pinsonneault} {et~al.}(2018){Pinsonneault}, {Elsworth}, {Tayar},
  {Serenelli}, {Stello}, {Zinn}, {Mathur}, {Garc{\'{\i}}a}, {Johnson},
  {Hekker}, {Huber}, {Kallinger}, {M{\'e}sz{\'a}ros}, {Mosser}, {Stassun},
  {Girardi}, {Rodrigues}, {Silva Aguirre}, {An}, {Basu}, {Chaplin}, {Corsaro},
  {Cunha}, {Garc{\'{\i}}a-Hern{\'a}ndez}, {Holtzman}, {J{\"o}nsson},
  {Shetrone}, {Smith}, {Sobeck}, {Stringfellow}, {Zamora}, {Beers},
  {Fern{\'a}ndez-Trincado}, {Frinchaboy}, {Hearty}, \&
  {Nitschelm}}]{Pinsonneault2018}
{Pinsonneault}, M.~H., {Elsworth}, Y.~P., {Tayar}, J., {et~al.} 2018, \apjs,
  239, 32

\bibitem[{{Riello} {et~al.}(2018){Riello}, {De Angeli}, {Evans}, {Busso},
  {Hambly}, {Davidson}, {Burgess}, {Montegriffo}, {Osborne}, {Kewley},
  {Carrasco}, {Fabricius}, {Jordi}, {Cacciari}, {van Leeuwen}, \&
  {Holland}}]{Riello2018}
{Riello}, M., {De Angeli}, F., {Evans}, D.~W., {et~al.} 2018, \aap, 616, A3

\bibitem[{{Riess} {et~al.}(2018){Riess}, {Casertano}, {Kenworthy}, {Scolnic},
  \& {Macri}}]{Riess2018}
{Riess}, A.~G., {Casertano}, S., {Kenworthy}, D., {Scolnic}, D., \& {Macri}, L.
  2018, ArXiv e-prints [\eprint[arXiv]{1810.03526}]

\bibitem[{{Rodrigues} {et~al.}(2017){Rodrigues}, {Bossini}, {Miglio},
  {Girardi}, {Montalb{\'a}n}, {Noels}, {Trabucchi}, {Coelho}, \&
  {Marigo}}]{Rodrigues2017}
{Rodrigues}, T.~S., {Bossini}, D., {Miglio}, A., {et~al.} 2017, \mnras, 467,
  1433

\bibitem[{{Rodrigues} {et~al.}(2014){Rodrigues}, {Girardi}, {Miglio},
  {Bossini}, {Bovy}, {Epstein}, {Pinsonneault}, {Stello}, {Zasowski}, {Prieto},
  {Chaplin}, {Hekker}, {Johnson}, {M{\'e}sz{\'a}ros}, {Mosser}, {Anders},
  {Basu}, {Beers}, {Chiappini}, {da Costa}, {Elsworth}, {Garc{\'{\i}}a},
  {P{\'e}rez}, {Hearty}, {Maia}, {Majewski}, {Mathur}, {Montalb{\'a}n},
  {Nidever}, {Santiago}, {Schultheis}, {Serenelli}, \&
  {Shetrone}}]{Rodrigues2014}
{Rodrigues}, T.~S., {Girardi}, L., {Miglio}, A., {et~al.} 2014, \mnras, 445,
  2758

\bibitem[{{Sahlholdt} {et~al.}(2018){Sahlholdt}, {Silva Aguirre}, {Casagrande},
  {Mosumgaard}, \& {Bojsen-Hansen}}]{Sahlholdt2018}
{Sahlholdt}, C.~L., {Silva Aguirre}, V., {Casagrande}, L., {Mosumgaard}, J.~R.,
  \& {Bojsen-Hansen}, M. 2018, \mnras, 476, 1931

\bibitem[{{Sharma} {et~al.}(2016){Sharma}, {Stello}, {Bland-Hawthorn}, {Huber},
  \& {Bedding}}]{Sharma2016}
{Sharma}, S., {Stello}, D., {Bland-Hawthorn}, J., {Huber}, D., \& {Bedding},
  T.~R. 2016, \apj, 822, 15

\bibitem[{{Silva Aguirre} {et~al.}(2012){Silva Aguirre}, {Casagrande}, {Basu},
  {Campante}, {Chaplin}, {Huber}, {Miglio}, {Serenelli}, {Ballot}, {Bedding},
  {Christensen-Dalsgaard}, {Creevey}, {Elsworth}, {Garc{\'{\i}}a}, {Gilliland},
  {Hekker}, {Kjeldsen}, {Mathur}, {Metcalfe}, {Monteiro}, {Mosser},
  {Pinsonneault}, {Stello}, {Weiss}, {Tenenbaum}, {Twicken}, \&
  {Uddin}}]{SilvaAguirre2012}
{Silva Aguirre}, V., {Casagrande}, L., {Basu}, S., {et~al.} 2012, \apj, 757, 99

\bibitem[{{Skrutskie} {et~al.}(2006){Skrutskie}, {Cutri}, {Stiening},
  {Weinberg}, {Schneider}, {Carpenter}, {Beichman}, {Capps}, {Chester},
  {Elias}, {Huchra}, {Liebert}, {Lonsdale}, {Monet}, {Price}, {Seitzer},
  {Jarrett}, {Kirkpatrick}, {Gizis}, {Howard}, {Evans}, {Fowler}, {Fullmer},
  {Hurt}, {Light}, {Kopan}, {Marsh}, {McCallon}, {Tam}, {Van Dyk}, \&
  {Wheelock}}]{Skrutskie2006}
{Skrutskie}, M.~F., {Cutri}, R.~M., {Stiening}, R., {et~al.} 2006, \aj, 131,
  1163

\bibitem[{{Stassun} \& {Torres}(2018)}]{Stassun2018}
{Stassun}, K.~G. \& {Torres}, G. 2018, \apj, 862, 61

\bibitem[{{Stello} {et~al.}(2009){Stello}, {Chaplin}, {Bruntt}, {Creevey},
  {Garc{\'{\i}}a-Hern{\'a}ndez}, {Monteiro}, {Moya}, {Quirion}, {Sousa},
  {Su{\'a}rez}, {Appourchaux}, {Arentoft}, {Ballot}, {Bedding},
  {Christensen-Dalsgaard}, {Elsworth}, {Fletcher}, {Garc{\'{\i}}a}, {Houdek},
  {Jim{\'e}nez-Reyes}, {Kjeldsen}, {New}, {R{\'e}gulo}, {Salabert}, \&
  {Toutain}}]{Stello2009}
{Stello}, D., {Chaplin}, W.~J., {Bruntt}, H., {et~al.} 2009, \apj, 700, 1589

\bibitem[{{Stello} {et~al.}(2015){Stello}, {Huber}, {Sharma}, {Johnson},
  {Lund}, {Handberg}, {Buzasi}, {Silva Aguirre}, {Chaplin}, {Miglio},
  {Pinsonneault}, {Basu}, {Bedding}, {Bland-Hawthorn}, {Casagrande}, {Davies},
  {Elsworth}, {Garcia}, {Mathur}, {Di Mauro}, {Mosser}, {Schneider},
  {Serenelli}, \& {Valentini}}]{Stello2015}
{Stello}, D., {Huber}, D., {Sharma}, S., {et~al.} 2015, \apjl, 809, L3

\bibitem[{{Stello} {et~al.}(2016){Stello}, {Vanderburg}, {Casagrande},
  {Gilliland}, {Silva Aguirre}, {Sandquist}, {Leiner}, {Mathieu}, \&
  {Soderblom}}]{Stello2016}
{Stello}, D., {Vanderburg}, A., {Casagrande}, L., {et~al.} 2016, \apj, 832, 133

\bibitem[{{Stello} {et~al.}(2017){Stello}, {Zinn}, {Elsworth}, {Garcia},
  {Kallinger}, {Mathur}, {Mosser}, {Sharma}, {Chaplin}, {Davies}, {Huber},
  {Jones}, {Miglio}, \& {Silva Aguirre}}]{Stello2017}
{Stello}, D., {Zinn}, J., {Elsworth}, Y., {et~al.} 2017, \apj, 835, 83

\bibitem[{{Tassoul}(1980)}]{Tassoul1980}
{Tassoul}, M. 1980, \apjs, 43, 469

\bibitem[{{Ulrich}(1986)}]{Ulrich1986}
{Ulrich}, R.~K. 1986, \apjl, 306, L37

\bibitem[{{Vandakurov}(1967)}]{Vandakurov1967}
{Vandakurov}, Y.~V. 1967, \azh, 44, 786

\bibitem[{{Vrard} {et~al.}(2015){Vrard}, {Mosser}, {Barban}, {Belkacem},
  {Elsworth}, {Kallinger}, {Hekker}, {Samadi}, \& {Beck}}]{Vrard2015}
{Vrard}, M., {Mosser}, B., {Barban}, C., {et~al.} 2015, \aap, 579, A84

\bibitem[{{White} {et~al.}(2011){White}, {Bedding}, {Stello},
  {Christensen-Dalsgaard}, {Huber}, \& {Kjeldsen}}]{White2011}
{White}, T.~R., {Bedding}, T.~R., {Stello}, D., {et~al.} 2011, \apj, 743, 161

\bibitem[{{Yu} {et~al.}(2018){Yu}, {Huber}, {Bedding}, {Stello}, {Hon},
  {Murphy}, \& {Khanna}}]{Yu2018}
{Yu}, J., {Huber}, D., {Bedding}, T.~R., {et~al.} 2018, \apjs, 236, 42

\bibitem[{{Zinn} {et~al.}(2018){Zinn}, {Pinsonneault}, {Huber}, \&
  {Stello}}]{Zinn2018}
{Zinn}, J.~C., {Pinsonneault}, M.~H., {Huber}, D., \& {Stello}, D. 2018, ArXiv
  e-prints [\eprint[arXiv]{1805.02650}]

\end{thebibliography}

\end{document}